\pdfoutput=1
\documentclass[journal]{IEEEtran}

\usepackage{cite}
\usepackage{authblk}
\usepackage{hyperref}
\usepackage{graphicx}
\graphicspath{{images/}}

\usepackage{booktabs}
\usepackage[font=small,labelfont=bf, font=bf]{caption}
\def\tablename{Table}
\renewcommand{\thetable}{\arabic{table}}
\usepackage{multirow}
\usepackage{subfig}

\usepackage{verbatim}

\usepackage{listings}

\usepackage{color}
\lstdefinestyle{mystyle}{
    basicstyle=\footnotesize,
    breakatwhitespace=false,         
    breaklines=true,                 
    captionpos=b,                    
    keepspaces=true,                 
    showspaces=false,                
    showstringspaces=false,
    showtabs=false,                  
    tabsize=2
}
\usepackage{amsmath}
\usepackage[colorinlistoftodos]{todonotes}
\usepackage{soul,xcolor}

\begin{document}
\newcommand{\etal}{~et~al.~}

\definecolor{dred}{rgb}{0.75,0.00, 0.00}
\definecolor{dgreen}{rgb}{0.00, 0.75, 0.00}
\definecolor{dblue}{rgb}{0.00, 0.00, 0.75}
\definecolor{stecol}{rgb}{0.75, 0.00, 0.75}
\newcommand{\gagan}[1]{[{\color{dgreen}GAGAN: #1}]}
\newcommand{\gaganFeedback}[1]{[{\color{dred}GAGAN: #1}]}

\newcommand{\gaganHighlight}[1]{[{\color{dred}#1}]}
\newcommand{\lorenzo}[1]{[{\color{red}LORENZO: #1}]}
\newcommand{\ahsan}[1]{[{\color{blue}Ahsan: #1}]}
\newcommand{\ste}[1]{[{\color{stecol}Stefano: #1}]}

\setstcolor{red}
\newcommand{\gaganRemove}[1] {\st{#1}}
%

\title{Near-Memory Computing: Past, Present, and Future}

\author{
  \IEEEauthorblockN{Gagandeep Singh\IEEEauthorrefmark{1}, Lorenzo Chelini\IEEEauthorrefmark{1}, Stefano Corda\IEEEauthorrefmark{1}, Ahsan Javed Awan\IEEEauthorrefmark{2}, Sander Stuijk\IEEEauthorrefmark{1}, Roel Jordans\IEEEauthorrefmark{1}, Henk Corporaal\IEEEauthorrefmark{1}, Albert-Jan Boonstra\IEEEauthorrefmark{3}}\\
    \IEEEauthorblockA{\IEEEauthorrefmark{1}Department of Electrical Engineering, Eindhoven University of Technology, Netherlands\\
    \{g.singh, l.chelini, s.corda, s.stuijk, r.jordans, h.corporaal\}@tue.nl}
    
    \IEEEauthorblockA{\IEEEauthorrefmark{2}Department of Cloud Systems and Platforms, Ericsson Research, Sweden
    \\ahsan.javed.awan@ericsson.com}
    
     \IEEEauthorblockA{\IEEEauthorrefmark{3}R\&D Department, Netherlands Institute for Radio Astronomy, ASTRON, Netherlands
    \\boonstra@astron.nl}
}

\maketitle

\begin{abstract}

The conventional approach of moving data to the CPU for computation has become a significant performance bottleneck for emerging scale-out data-intensive applications due to their limited data reuse. At the same time, the advancement in 3D integration technologies has made the decade-old concept of coupling compute units close to the memory --- called near-memory computing (NMC) --- more viable. Processing right at the ``home'' of data can significantly diminish the data movement problem of data-intensive applications.

In this paper, we survey the prior art on NMC across various dimensions (architecture, applications, tools, etc.) and identify the key challenges and open issues with future research directions. We also provide a glimpse of our approach to near-memory computing that includes i) NMC specific microarchitecture independent application characterization ii) a compiler framework to offload the NMC kernels on our target NMC platform and iii) an analytical model to evaluate the potential of NMC.



\end{abstract}


%
\IEEEpeerreviewmaketitle

\section{Introduction}
\label{sec:introduction}
    Over the years, memory technology has not been able to keep up with advancements in processor technology in terms of latency and energy consumption, which is referred to as the \textit{memory wall}~\cite{Wulf1995}. Earlier, system architects tried to bridge this gap by introducing memory hierarchies that mitigated some of the disadvantages of off-chip DRAMs. However, the limited number of pins on the memory package is not able to meet today's bandwidth demands of multicore processors. Furthermore, with the demise of Dennard scaling~\cite{1050511}, slowing of Moore's law, and dark silicon computer performance has reached a plateau~\cite{6175879}. 

At the same time, we are witnessing an enormous amount of data being generated across multiple areas like radio astronomy, material science, chemistry, health sciences etc~\cite{nair2015active}. In radio astronomy, for example, the first phase of the Square Kilometre Array (SKA) aims at processing over 100 terabytes of raw data samples per second, yielding of the order of 300 petabytes of SKA data products annually~\cite{6898703}. The SKA currently is in the design phase with anticipated construction in South Africa and Australia in the first half of the coming decade. These radio-astronomy applications usually exhibit massive data parallelism and low operational intensity with a limited locality. On traditional systems, these applications cause frequent data movement between the memory subsystem and the processor that has a severe impact on performance and energy efficiency. Likewise, cluster computing frameworks like Apache Spark that enable distributed in-memory processing of batch and streaming data also exhibit those mentioned above behaviour~\cite{awan2016micro,javed2015performance,awan2016node}. Therefore, a lot of current research is being focused on coming up with innovative manufacturing technologies and architectures to overcome these problems. 

Today's memory hierarchy usually consists of multiple levels of cache, the main memory, and storage. The traditional approach is to move data up to caches from the storage and then process it. In contrast, near-memory computing (NMC) aims at processing close to where the data resides. This data-centric approach couples compute units close to the data and seek to minimize the expensive data movements. Notably, three-dimensional stacking is touted as the true enabler of processing close to the memory. It allows the stacking of logic and memory together using through-silicon via's (TSVs) that helps in reducing memory access latency, power consumption and provides much higher bandwidth~\cite{6242474}. Micron's Hybrid Memory Cube (HMC)~\cite{7477494}, High Bandwidth Memory (HBM)~\cite{6757501} from AMD and Hynix, and Samsung's Wide I/O~\cite{6025219} are the competing products in the 3D memory arena. 

\begin{figure*}[t]
\centering
\includegraphics[width=1\textwidth]{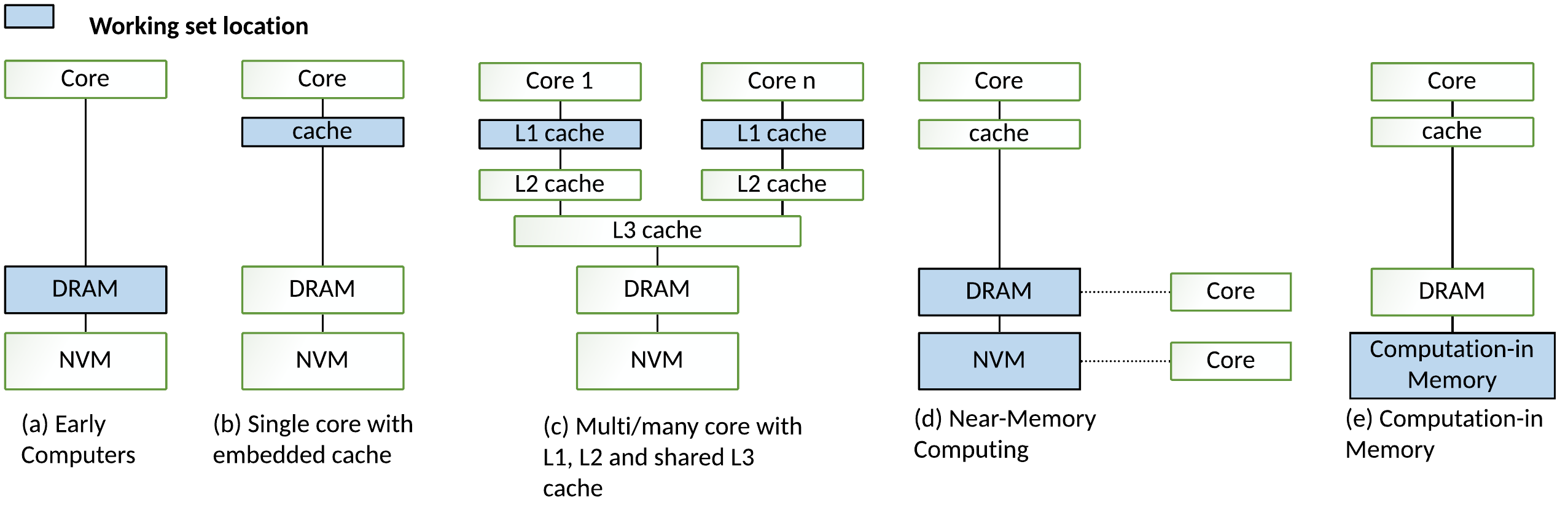}
\caption{Classification of computing systems based on working set location, which is referred to as a working set~\cite{7092668}. Prior systems were based on a CPU-centric approach where data is moved to the core for processing (Figure 1 (a)-(c)), whereas now with near-memory processing (Figure 1 (d)) the processing cores are brought to the place where data resides. Computation-in-memory (Figure 1 (e)) further reduces data movement by using memories with compute capability (e.g. memristors, phase change memory).
\label{fig:allsystems}}
\end{figure*}

\begin{figure}[!hb]
\centering
\includegraphics[width=0.9\linewidth]{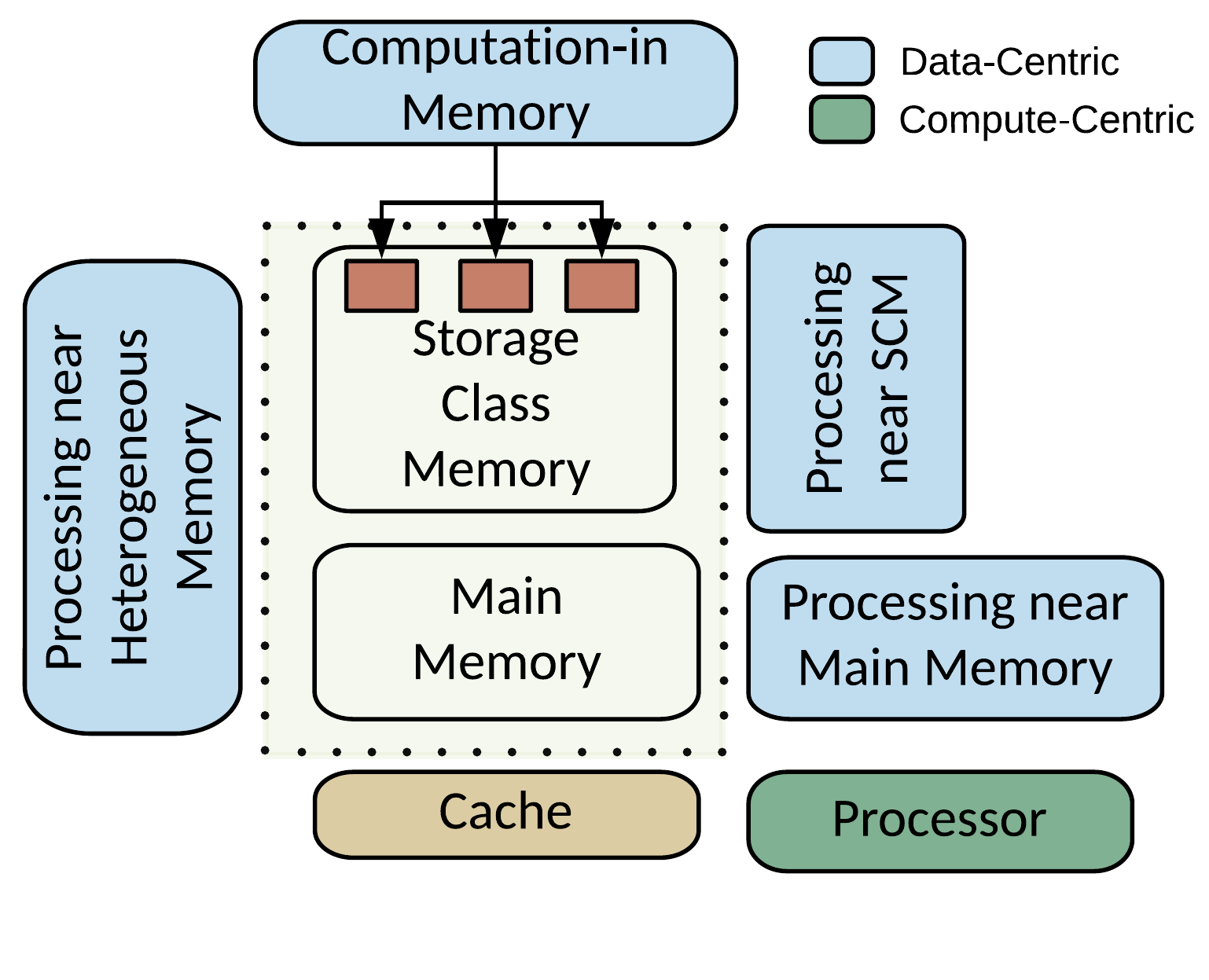}
\caption{Processing options in the memory hierarchy highlighting the conventional compute-centric and the modern data-centric approach
\label{fig:taxonomy}}
\end{figure}

Figure~\ref{fig:allsystems} depicts the system evolution based on the information referenced by a program during execution, which is referred to as a working set~\cite{7092668}. Prior systems were based on a CPU-centric approach where data is moved to the core for processing (Figure 1 (a)-(c)), whereas now with near-memory processing (Figure 1 (d)) the processing cores are brought to the place where data resides. Computation-in-memory (Figure 1 (e)) paradigm aims at reducing data movement completely by using memories with compute capability (e.g. memristors, phase-change memory (PCM)).

The paper aims to analyze and organize the extensive body of literature related to the novel area of near-memory computing. Figure~\ref{fig:taxonomy} shows a high-level view of our classification that is based on the level in the memory hierarchy and further split into the type of compute implementation (programmable, fixed-function, or reconfigurable). Conceptually the approach  of near-memory computing can be applied to any level or type of memory to improve the overall system performance. In our taxonomy, we do not include magnetic disk-based systems as they can no longer offer timely response due to the high access latency and a high failure rate of disks~\cite{7097722}. Nevertheless, there have been research efforts towards providing processing capabilities in the disk. However, it was not adopted widely by the industry due to the marginal performance improvement that could not justify the associated cost~\cite{Keeton:1998:CID:290593.290602,Riedel:1998:ASL:645924.671345}. Instead, we include emerging non-volatile memories termed as storage-class memories (SCM)~\cite{7151782} which are trying to fill the latency gap between DRAMs and disks. Building upon similar efforts~\cite{siegl2016data, ghose2018enabling}, we make the following contributions:

\begin{itemize}

\item We analyze and organize the body of literature on near-memory computing under various dimensions.
\item We provide guidelines for design space exploration.
\item We present our near-memory system outlining application characterization and compiler framework. Also, we include an analytic model to illustrate the potential of near-memory computing for memory intensive application with a comparison to the current CPU-centric approach.
\item We outline the directions for future research and highlight current challenges.

\end{itemize}

This work extends our previous work~\cite{629e227db47e4e89b539b0e072ecce08} by including our approach to near-memory computing (Section~\ref{sec:system}).  This outlines our application characterization and compilation framework. 
Furthermore, we evaluate more recent work in the NMC domain and provide new insights and research directions. The remainder of this article is structured as follows: Section~\ref{sec:background} provides a historical overview of near-memory computing and related work. Section~\ref{sec:classificationmodel} outlines the evaluation and classification scheme for NMC at main memory (Section~\ref{sec:procmainmem}) and storage class memory (Section~\ref{sec:procstorage}).  Section~\ref{sec:interondpcachevirtual} highlights the challenges with cache coherence, virtual memory, the lack of programming models, and data mapping schemes for NMC. In Section~\ref{sec:modelingmethod}, we look into the tools and techniques used to perform the design space exploration for these systems. This section also illustrates the importance of application characterization. Section~\ref{sec:system} presents our approach to near-memory computing. Additionally, we include a high-level analytic approach with a comparison to a traditional CPU-centric approach. Finally, Section~\ref{sec:futuredirections} highlights the lessons learned and future research directions.

\begin{table}[ht]
\renewcommand{\arraystretch}{1.2}
\centering
\resizebox{\columnwidth}{!}{
\begin{tabular}{l|l|l|l}
\hline
\textbf{} & \textbf{Property} & \textbf{Abbrev} & \textbf{Description} \\ \hline
\multirow{11}{*}{\textbf{Memory}} & \multirow{3}{*}{Hierarchy} & MM & Main memory \\
 &  & SCM & Storage class memory \\
 &  & HM & Heterogenous memory \\ \cline{2-4} 
 & \multirow{6}{*}{Type} & C3D & Commercial 3D memory \\
 &  & DIMM & Dual in-line memory module \\
 &  & PCM & Phase change memory \\
 &  & DRAM & Dynamic random-access memory \\
 &  & SSD & Flash SSD memory \\
 &  & LRDIMM & Load-reduce DIMM \\ \cline{2-4} 
 & \multirow{2}{*}{Integration} & US & Conventional unstacked \\
 &  & S & Stacked using 2.5D or 3D \\ \hline
\multirow{12}{*}{\textbf{Processing}} & \multirow{5}{*}{\begin{tabular}[c]{@{}l@{}}NMC/Host\\ Unit\end{tabular}} & CPU & Central processing unit \\
 &  & GPU & Graphics processing unit \\
 &  & APU & Accelerated processing unit \\
 &  & FPGA & Field programmable gate array \\
 &  & CGRA & \begin{tabular}[c]{@{}l@{}}Coarse-grained reconfigurable \\ architecture\end{tabular} \\
 &  & ACC & Application specific accelerator \\ \cline{2-4} 
 & \multirow{3}{*}{\begin{tabular}[c]{@{}l@{}}Implemen-\\ tation\end{tabular}} & P & Programmable unit \\
 &  & F & Fixed function unit \\
 &  & R & Reconfigurable unit \\ \cline{2-4} 
 & \multirow{3}{*}{Granularity} & I & Instruction \\
 &  & K & Kernel \\
 &  & A & Application \\ \cline{2-4} 
 & Host Unit &  & Type of host unit \\ \hline
\multirow{3}{*}{\textbf{Tool}} & \multirow{3}{*}{\begin{tabular}[c]{@{}l@{}}Evaluation \\ Technique\end{tabular}} & A & Analytic \\
 &  & S & Simulation \\
 &  & P & Prototype/Hardware \\ \hline
\multirow{3}{*}{\textbf{\begin{tabular}[c]{@{}l@{}}Interop-\\ erability\end{tabular}}} & \begin{tabular}[c]{@{}l@{}}Programming \\ Model\end{tabular} & - & \begin{tabular}[c]{@{}l@{}}Programming model used \\ for the accelerator\end{tabular} \\
 & \begin{tabular}[c]{@{}l@{}}Cache \\ Coherence\end{tabular} & Y/N & Mechanism for cache coherence \\
 & \begin{tabular}[c]{@{}l@{}}Virtual \\ Memory\end{tabular} & Y/N & Virtual memory support \\ \hline
\textbf{\begin{tabular}[c]{@{}l@{}}App. \\ Domain\end{tabular}} & Workload & - & \begin{tabular}[c]{@{}l@{}}Target application domain \\ for the architecture\end{tabular} \\ \hline
\end{tabular}
}
\caption{Classification table, and legend for table~\ref{tab:arcClassification}}
\label{tab:arcLegend}
\end{table}
\section{Background and Related work}
\label{sec:background}
The idea of processing close to the memory dates back to the 1960s~\cite{ALogicinMemoryComputer}; however, the first appearance of NMC systems can be traced back to the early 1990s~\cite{5727436,Kogge1994,fbram1994,375174,592312,808425}. An example was \textit{Vector IRAM} (VIRAM)~\cite{612252}, where the researchers developed a vector processor with an on-chip embedded DRAM (eDRAM) to exploit data parallelism in multimedia applications. Although promising results were obtained, these NMC systems did not penetrate the market, and their adoption remained limited. One of the main reasons was attributed to technological limitations. Mainly because the amount of on-chip memory they could integrate with the vector processor was limited due to the difference in logic and memory technology processes.

\begin{figure}[]
\centering
\includegraphics[scale=0.14]{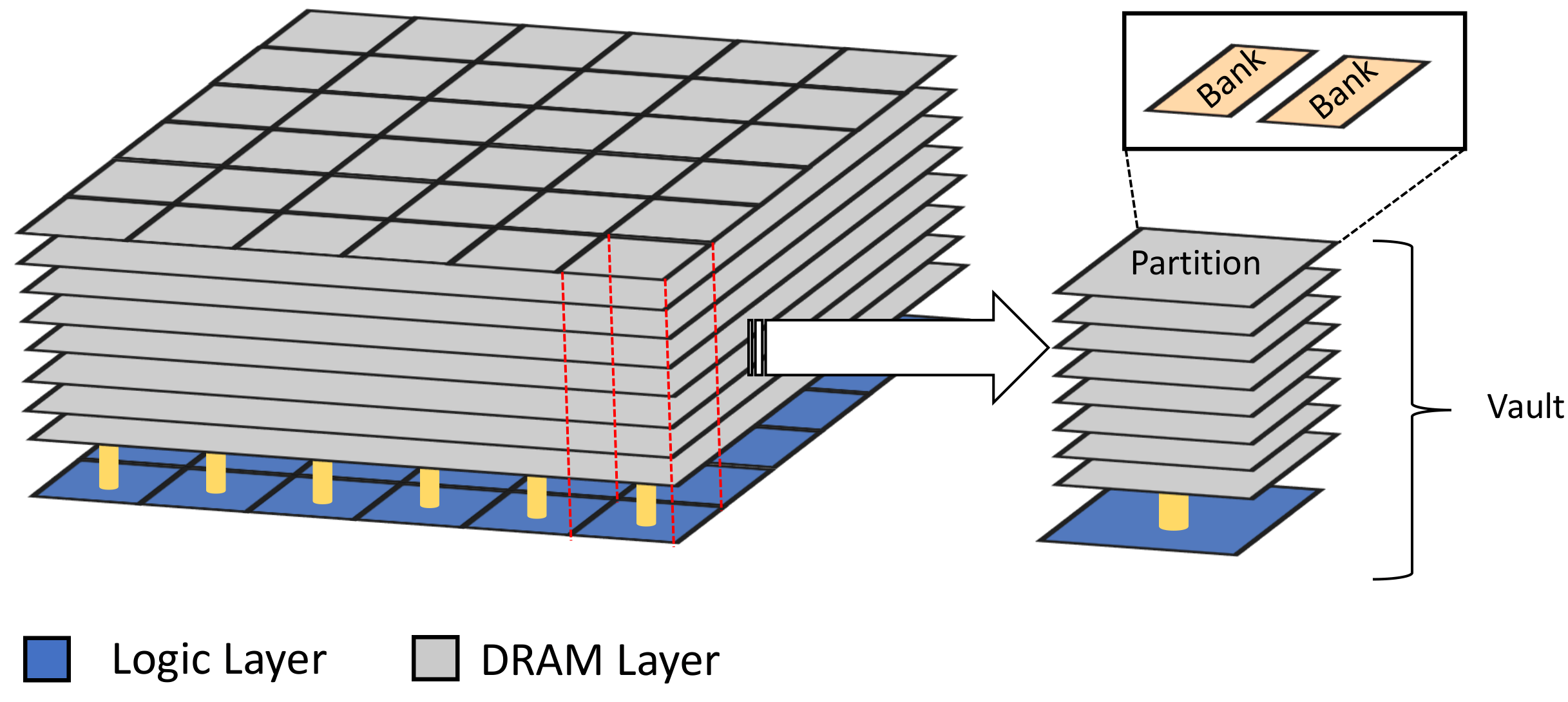}
\caption{Micron's Hybrid Memory Cube (HMC)~\cite{7477494} comprising DRAM layers stacked on top of a logic layer via through silicon via (TSV). The memory organization is divided into vaults with each vault consisting of multiple DRAM banks. 
\label{fig:HMCLayout}}
\end{figure}

Today, after almost two decades of dormancy, research in NMC is regaining attention. This resurgence is largely attributed to following three reasons. First, technological advancements in 3D (see Figure~\ref{fig:HMCLayout}) and 2.5D stacking that blends logic and memory in the same package. Second, moving the computation closer to where the data reside allows for sidestepping the performance and energy bottlenecks due to data movement by circumventing memory-package pin-count limitations. Third, with the advent of modern data-intensive applications in areas like material science, astronomy, health care, etc., calls for newer architectures. As a result, researches have proposed various NMC designs and proved their potential in enhancing performance in many applications~\cite{7284059, hsieh2016accelerating, Azarkhish:2016:DEP:2963802.2963805, 7927081}.

In the literature, NMC has manifested with names such as \textit{processing-in memory} (PIM),  \textit{near-data processing} (NDP), \textit{near-memory processing} (NMP), or in case of non volatile memories as \textit{in-storage processing} (ISP). However, all these terms fall under the same umbrella of \textit{near-memory computing} (NMC) with the core principle of doing processing close to the memory. With the arrival of true \textit{in-situ} computing called \textit{computation-in-memory} through novel devices such as memristors and phase-change memory, it would be prudent to merge all these synonyms mentioned above for the same concept under the umbrella of \textit{near-memory computing}. This unification provides systematization and positioning of new proposals within the existing works.

Loh et al.~\cite{loh2013processing} in their position paper presented an initial taxonomy for processing in memory. It is based on computing interface with software and a separate division for transparent software features. Siegl et al.~\cite{siegl2016data} in an overview paper gave a historical evolution of NMC. However, their classification metrics are not systematic and does not highlight current challenges and future research directions in this domain. Similar to our paper, Ghose\etal\cite{ghose2018enabling} in their most recent work provide a thorough overview of the mechanisms and challenges in the field of near-memory computing. Unlike us, however, they don't focus on providing systematization to the literature. In our review, we characterized near-memory computing literature in various dimensions starting from the memory level where the paradigm of near-memory processing is applied, to the type of processing unit, memory integration, and kind of workloads/applications.

\begin{table*}[ht]
\renewcommand{\arraystretch}{1.2}
\centering
\tabcolsep=0.22cm

\resizebox{\textwidth}{!}{

\begin{tabular}{@{}cccclccllclllcl@{}|}
\hline
\multicolumn{2}{|c}{\textbf{NMC Architecture}}& \multicolumn{3}{|c|}{\textbf{Memory}}& \multicolumn{4}{c|}{\textbf{Processing}} &\textbf{Tool}&\multicolumn{3}{|c|}{\textbf{Interoperability}}& \multicolumn{1}{|c|}{\textbf{App. Domain}} \\ \hline
\multicolumn{1}{|l}{\rotatebox[origin=c]{90}{\textbf{Architecture}}} & \multicolumn{1}{c|}{\rotatebox[origin=c]{90}{\textbf{Year}}}&
\multicolumn{1}{c|}{\rotatebox[origin=c]{90}{\textbf{Hierarchy}}}   & \multicolumn{1}{c|}{\rotatebox[origin=c]{90}{\textbf{Type}}} & \multicolumn{1}{l|}{\rotatebox[origin=c]{90}{\textbf{Integration}}} & \multicolumn{1}{c|}{\rotatebox[origin=c]{90}{\textbf{NMC Unit}}} & \multicolumn{1}{l|}{\rotatebox[origin=c]{90}{\textbf{Implementation}}} &
\multicolumn{1}{c|}{\rotatebox[origin=c]{90}{\textbf{Granularity}}} & \multicolumn{1}{l|}{\rotatebox[origin=c]{90}{\textbf{Host Unit}}} & \multicolumn{1}{c|}{\rotatebox[origin=c]{90}{\textbf{Evaluation}}} & 
\multicolumn{1}{c|}{\rotatebox[origin=c]{90}{\textbf{Programming Model} }} &
\multicolumn{1}{l|}{\rotatebox[origin=c]{90}{\textbf{Cache Coherence}}} & \multicolumn{1}{l|}{\rotatebox[origin=c]{90}{\textbf{Virtual Memory}}} &
\multicolumn{1}{c|}{\rotatebox[origin=c]{90}{\textbf{Workload}}}\\ 

\multicolumn{1}{|c}{}& 
\multicolumn{1}{c|}{}& 
\multicolumn{1}{c|}{}& 
\multicolumn{1}{c|}{}&  
\multicolumn{1}{l|}{}& 
\multicolumn{1}{c|}{}& 
\multicolumn{1}{l|}{}& 
\multicolumn{1}{l|}{}& 
\multicolumn{1}{l|}{}& 
\multicolumn{1}{l|}{}& 
\multicolumn{1}{l|}{}&
\multicolumn{1}{l|}{}&  
\multicolumn{1}{l|}{}&  
\multicolumn{1}{l|}{}\\ \hline 

\multicolumn{1}{|l}{XSD~\cite{cho2013xsd}}&
\multicolumn{1}{c|}{2013} &
\multicolumn{1}{c|}{SCM} &
\multicolumn{1}{c|}{SSD} & 
\multicolumn{1}{l|}{US}  & 
\multicolumn{1}{c|}{GPU}&
\multicolumn{1}{l|}{P} &
\multicolumn{1}{l|}{A} & 
\multicolumn{1}{l|}{CPU} &
\multicolumn{1}{c|}{S}&
\multicolumn{1}{l|}{MapReduce}&
\multicolumn{1}{l|}{-}& 
\multicolumn{1}{l|}{-} & 
\multicolumn{1}{l|}{MapReduce Workloads}\\ \hline

\multicolumn{1}{|l}{SmartSSD~\cite{kang2013enabling}}& 
\multicolumn{1}{c|}{2013} & 
\multicolumn{1}{c|}{SCM} & 
\multicolumn{1}{c|}{SSD} &  
\multicolumn{1}{l|}{US}  & 
\multicolumn{1}{c|}{CPU}& 
\multicolumn{1}{l|}{P} & 
\multicolumn{1}{l|}{A} & 
\multicolumn{1}{l|}{CPU} & 
\multicolumn{1}{c|}{P}& 
\multicolumn{1}{l|}{MapReduce}&
\multicolumn{1}{l|}{Y}&  
\multicolumn{1}{l|}{N} &  
\multicolumn{1}{l|}{Database}\\ \hline 

\multicolumn{1}{|l}{WILLOW~\cite{seshadri2014willow}}& 
\multicolumn{1}{c|}{2014} & 
\multicolumn{1}{c|}{SCM} & 
\multicolumn{1}{c|}{SSD} &  
\multicolumn{1}{l|}{US}  & 
\multicolumn{1}{c|}{CPU}& 
\multicolumn{1}{l|}{P} & 
\multicolumn{1}{l|}{K} & 
\multicolumn{1}{l|}{CPU} & 
\multicolumn{1}{c|}{P}& 
\multicolumn{1}{l|}{API}&
\multicolumn{1}{l|}{Y}&  
\multicolumn{1}{l|}{-}&  
\multicolumn{1}{l|}{Generic}\\ \hline 

\multicolumn{1}{|l}{NDC~\cite{6844483}}&
\multicolumn{1}{c|}{2014} &
\multicolumn{1}{c|}{MM} &
\multicolumn{1}{c|}{C3D} & 
\multicolumn{1}{l|}{S}  & 
\multicolumn{1}{c|}{CPU}&
\multicolumn{1}{l|}{P} &
\multicolumn{1}{l|}{K} & 
\multicolumn{1}{l|}{CPU} &
\multicolumn{1}{c|}{S}&
\multicolumn{1}{l|}{MapReduce}&
\multicolumn{1}{l|}{R} & 
\multicolumn{1}{l|}{N}& 
\multicolumn{1}{l|}{MapReduce Workloads}\\ \hline

\multicolumn{1}{|l}{TOP-PIM~\cite{zhang2014top}}& 
\multicolumn{1}{c|}{2014} & 
\multicolumn{1}{c|}{MM} & 
\multicolumn{1}{c|}{C3D} &  
\multicolumn{1}{l|}{S}  & 
\multicolumn{1}{c|}{APU}& 
\multicolumn{1}{l|}{P} & 
\multicolumn{1}{l|}{K} & 
\multicolumn{1}{l|}{CPU} & 
\multicolumn{1}{c|}{S}& 
\multicolumn{1}{l|}{OpenCL}&
\multicolumn{1}{l|}{Y}&  
\multicolumn{1}{l|}{-} &  
\multicolumn{1}{l|}{Graph and HPC}\\ \hline 

\multicolumn{1}{|l}{AMC~\cite{nair2015active}}&
\multicolumn{1}{c|}{2015} &
\multicolumn{1}{c|}{MM} &
\multicolumn{1}{c|}{C3D} & 
\multicolumn{1}{l|}{S}  & 
\multicolumn{1}{c|}{CPU}&
\multicolumn{1}{l|}{P} &
\multicolumn{1}{l|}{K} & 
\multicolumn{1}{l|}{CPU} &
\multicolumn{1}{c|}{S}&
\multicolumn{1}{l|}{OpenMP}&
\multicolumn{1}{l|}{Y}& 
\multicolumn{1}{l|}{Y} & 
\multicolumn{1}{l|}{HPC}\\ \hline

\multicolumn{1}{|l}{JAFAR~\cite{xi2015beyond}}&
\multicolumn{1}{c|}{2015} &
\multicolumn{1}{c|}{MM} &
\multicolumn{1}{c|}{DIMM} & 
\multicolumn{1}{l|}{US}  & 
\multicolumn{1}{c|}{ACC}&
\multicolumn{1}{l|}{F} &
\multicolumn{1}{l|}{K} & 
\multicolumn{1}{l|}{CPU} &
\multicolumn{1}{c|}{S}&
\multicolumn{1}{l|}{API}&
\multicolumn{1}{l|}{-}& 
\multicolumn{1}{l|}{Y}& 
\multicolumn{1}{l|}{Database}\\ \hline

\multicolumn{1}{|l}{TESSERACT~\cite{7284059}}& 
\multicolumn{1}{c|}{2015} & 
\multicolumn{1}{c|}{MM} & 
\multicolumn{1}{c|}{C3D} &  
\multicolumn{1}{l|}{S}  & 
\multicolumn{1}{c|}{CPU}& 
\multicolumn{1}{l|}{F} & 
\multicolumn{1}{l|}{A} & 
\multicolumn{1}{l|}{CPU} & 
\multicolumn{1}{c|}{S}& 
\multicolumn{1}{l|}{API}&
\multicolumn{1}{l|}{Y}&  
\multicolumn{1}{l|}{N} &  
\multicolumn{1}{l|}{Graph processing}\\ \hline 

\multicolumn{1}{|l}{Gokhale~\cite{gokhale2015near}}&
\multicolumn{1}{c|}{2015} &
\multicolumn{1}{c|}{MM} &
\multicolumn{1}{c|}{C3D} & 
\multicolumn{1}{l|}{S}  & 
\multicolumn{1}{c|}{ACC}&
\multicolumn{1}{l|}{F} &
\multicolumn{1}{l|}{K} & 
\multicolumn{1}{l|}{CPU} &
\multicolumn{1}{c|}{S}&
\multicolumn{1}{l|}{API}&
\multicolumn{1}{l|}{Y}& 
\multicolumn{1}{l|}{Y} & 
\multicolumn{1}{l|}{Generic}\\ \hline

\multicolumn{1}{|l}{HRL~\cite{7446059}}&
\multicolumn{1}{c|}{2015} &
\multicolumn{1}{c|}{MM} &
\multicolumn{1}{c|}{C3D} & 
\multicolumn{1}{l|}{S}  & 
\multicolumn{1}{c|}{CGRA+FPGA}& 
\multicolumn{1}{l|}{R} & 
\multicolumn{1}{l|}{A} & 
\multicolumn{1}{l|}{CPU} &
\multicolumn{1}{c|}{S}&
\multicolumn{1}{l|}{MapReduce}& 
\multicolumn{1}{l|}{Y}& 
\multicolumn{1}{l|}{N} & 
\multicolumn{1}{l|}{Data analytics}\\ \hline

\multicolumn{1}{|l}{ProPRAM~\cite{wang2015propram}}& 
\multicolumn{1}{c|}{2015} & 
\multicolumn{1}{c|}{SCM} & 
\multicolumn{1}{c|}{PCM} &  
\multicolumn{1}{l|}{US}  & 
\multicolumn{1}{c|}{CPU}& 
\multicolumn{1}{l|}{P} & 
\multicolumn{1}{l|}{I} & 
\multicolumn{1}{l|}{-} & 
\multicolumn{1}{c|}{S}& 
\multicolumn{1}{l|}{ISA Extension}&
\multicolumn{1}{l|}{-}&  
\multicolumn{1}{l|}{-} &  
\multicolumn{1}{l|}{Data analytics}\\ \hline 

\multicolumn{1}{|l}{BlueDBM~\cite{Jun:2015:BAB:2872887.2750412}}& 
\multicolumn{1}{c|}{2015} & 
\multicolumn{1}{c|}{SCM} & 
\multicolumn{1}{c|}{SSD} &  
\multicolumn{1}{l|}{US}  & 
\multicolumn{1}{c|}{FPGA}& 
\multicolumn{1}{l|}{R} & 
\multicolumn{1}{l|}{K} & 
\multicolumn{1}{l|}{-} & 
\multicolumn{1}{c|}{P}& 
\multicolumn{1}{l|}{API}&
\multicolumn{1}{l|}{-}&  
\multicolumn{1}{l|}{-} &  
\multicolumn{1}{l|}{Data analytics} \\ \hline 


\multicolumn{1}{|l}{NDA~\cite{7056040}}& 
\multicolumn{1}{c|}{2015} & 
\multicolumn{1}{c|}{MM} & 
\multicolumn{1}{c|}{LRDIMM} &  
\multicolumn{1}{l|}{S}  & 
\multicolumn{1}{c|}{CGRA}& 
\multicolumn{1}{l|}{R} & 
\multicolumn{1}{l|}{K} & 
\multicolumn{1}{l|}{CPU} & 
\multicolumn{1}{c|}{S}& 
\multicolumn{1}{l|}{OpenCL}&
\multicolumn{1}{l|}{Y}&  
\multicolumn{1}{l|}{Y} &  
\multicolumn{1}{l|}{MapReduce Workloads}\\ \hline 

\multicolumn{1}{|l}{PIM-enabled~\cite{ahn2015pim}}& 
\multicolumn{1}{c|}{2015} & 
\multicolumn{1}{c|}{MM} & 
\multicolumn{1}{c|}{C3D} &  
\multicolumn{1}{l|}{S}  & 
\multicolumn{1}{c|}{ACC}& 
\multicolumn{1}{l|}{F} & 
\multicolumn{1}{l|}{I} & 
\multicolumn{1}{l|}{CPU} & 
\multicolumn{1}{c|}{S}& 
\multicolumn{1}{l|}{ISA extension}&
\multicolumn{1}{l|}{Y}&  
\multicolumn{1}{l|}{Y} &  
\multicolumn{1}{l|}{Generic}\\ \hline 


\multicolumn{1}{|l}{IMPICA~\cite{hsieh2016accelerating}}& 
\multicolumn{1}{c|}{2016} & 
\multicolumn{1}{c|}{MM} & 
\multicolumn{1}{c|}{C3D} &  
\multicolumn{1}{l|}{S}  & 
\multicolumn{1}{c|}{ACC}& 
\multicolumn{1}{l|}{F} & 
\multicolumn{1}{l|}{K} & 
\multicolumn{1}{l|}{CPU} & 
\multicolumn{1}{c|}{S}& 
\multicolumn{1}{l|}{API}&
\multicolumn{1}{l|}{Y}&  
\multicolumn{1}{l|}{Y} &  
\multicolumn{1}{l|}{Pointer chasing}\\ \hline 

\multicolumn{1}{|l}{TOM~\cite{7551394}}&
\multicolumn{1}{c|}{2016} &
\multicolumn{1}{c|}{MM} &
\multicolumn{1}{c|}{C3D} & 
\multicolumn{1}{l|}{S}  & 
\multicolumn{1}{c|}{GPU}&
\multicolumn{1}{l|}{P} &
\multicolumn{1}{l|}{K} & 
\multicolumn{1}{l|}{GPU} &
\multicolumn{1}{c|}{S}&
\multicolumn{1}{l|}{CUDA}&
\multicolumn{1}{l|}{Y}& 
\multicolumn{1}{l|}{Y} & 
\multicolumn{1}{l|}{Generic}\\ \hline


\multicolumn{1}{|l}{BISCUIT~\cite{gu2016biscuit}}& 
\multicolumn{1}{c|}{2016} & 
\multicolumn{1}{c|}{SCM} & 
\multicolumn{1}{c|}{SSD} &  
\multicolumn{1}{l|}{US}  & 
\multicolumn{1}{c|}{ACC}& 
\multicolumn{1}{l|}{F} & 
\multicolumn{1}{l|}{K} & 
\multicolumn{1}{l|}{CPU} & 
\multicolumn{1}{c|}{P}& 
\multicolumn{1}{l|}{API}&
\multicolumn{1}{l|}{-}&  
\multicolumn{1}{l|}{-} &  
\multicolumn{1}{l|}{Database}\\ \hline 

\multicolumn{1}{|l}{Pattnaik~\cite{7756764}}& 
\multicolumn{1}{c|}{2016} & 
\multicolumn{1}{c|}{MM} & 
\multicolumn{1}{c|}{C3D} &  
\multicolumn{1}{l|}{S}  & 
\multicolumn{1}{c|}{GPU}& 
\multicolumn{1}{l|}{P} & 
\multicolumn{1}{l|}{K} & 
\multicolumn{1}{l|}{GPU} & 
\multicolumn{1}{c|}{S}& 
\multicolumn{1}{l|}{CUDA}&
\multicolumn{1}{l|}{Y}&  
\multicolumn{1}{l|}{-} &  
\multicolumn{1}{l|}{Generic}\\ \hline 

\multicolumn{1}{|l}{CARIBOU~\cite{istvan2017caribou}}& 
\multicolumn{1}{c|}{2017} & 
\multicolumn{1}{c|}{SCM} & 
\multicolumn{1}{c|}{DRAM} &  
\multicolumn{1}{l|}{US}  & 
\multicolumn{1}{c|}{FPGA}& 
\multicolumn{1}{l|}{R} & 
\multicolumn{1}{l|}{K} & 
\multicolumn{1}{l|}{CPU} & 
\multicolumn{1}{c|}{P}& 
\multicolumn{1}{l|}{API}&
\multicolumn{1}{l|}{-}&  
\multicolumn{1}{l|}{-} &  
\multicolumn{1}{l|}{Database}\\ \hline 

\multicolumn{1}{|l}{Vermij~\cite{vermij2017sorting}}&
\multicolumn{1}{c|}{2017} &
\multicolumn{1}{c|}{MM} &
\multicolumn{1}{c|}{C3D} & 
\multicolumn{1}{l|}{S}  & 
\multicolumn{1}{c|}{ACC}&
\multicolumn{1}{l|}{F} &
\multicolumn{1}{l|}{A} & 
\multicolumn{1}{l|}{CPU} &
\multicolumn{1}{c|}{S}&
\multicolumn{1}{l|}{API}&
\multicolumn{1}{l|}{Y}& 
\multicolumn{1}{l|}{Y} & 
\multicolumn{1}{l|}{Sorting}\\ \hline



\multicolumn{1}{|l}{SUMMARIZER~\cite{koo2017summarizer}}& 
\multicolumn{1}{c|}{2017} & 
\multicolumn{1}{c|}{SCM} & 
\multicolumn{1}{c|}{SSD} &  
\multicolumn{1}{l|}{US}  & 
\multicolumn{1}{c|}{CPU}& 
\multicolumn{1}{l|}{P} & 
\multicolumn{1}{l|}{K} & 
\multicolumn{1}{l|}{CPU} & 
\multicolumn{1}{c|}{P}& 
\multicolumn{1}{l|}{API}&
\multicolumn{1}{l|}{-}&  
\multicolumn{1}{l|}{-} &  
\multicolumn{1}{l|}{Database}\\ \hline 

\multicolumn{1}{|l}{MONDRIAN~\cite{de2017mondrian}}& 
\multicolumn{1}{c|}{2017} & 
\multicolumn{1}{c|}{MM} & 
\multicolumn{1}{c|}{C3D} &  
\multicolumn{1}{l|}{S}  & 
\multicolumn{1}{c|}{CPU}& 
\multicolumn{1}{l|}{P} & 
\multicolumn{1}{l|}{K} & 
\multicolumn{1}{l|}{CPU} & 
\multicolumn{1}{c|}{A+S}& 
\multicolumn{1}{l|}{API}&
\multicolumn{1}{l|}{-}&  
\multicolumn{1}{l|}{Y} &  
\multicolumn{1}{l|}{Data analytics}\\ \hline 

\multicolumn{1}{|l}{GraphPIM~\cite{nai2017graphpim}}&
\multicolumn{1}{c|}{2017} &
\multicolumn{1}{c|}{MM} &
\multicolumn{1}{c|}{DRAM} & 
\multicolumn{1}{l|}{US}  & 
\multicolumn{1}{c|}{ACC}&
\multicolumn{1}{l|}{F} &
\multicolumn{1}{l|}{I} & 
\multicolumn{1}{l|}{CPU} &
\multicolumn{1}{c|}{S}&
\multicolumn{1}{l|}{API}&
\multicolumn{1}{l|}{Y}& 
\multicolumn{1}{l|}{N} & 
\multicolumn{1}{l|}{Graph}\\ \hline

\multicolumn{1}{|l}{MCN~\cite{alian2018application}}&
\multicolumn{1}{c|}{2018} &
\multicolumn{1}{c|}{MM} &
\multicolumn{1}{c|}{DRAM} & 
\multicolumn{1}{l|}{US}  & 
\multicolumn{1}{c|}{CPU}&
\multicolumn{1}{l|}{P} &
\multicolumn{1}{l|}{K} & 
\multicolumn{1}{l|}{CPU} &
\multicolumn{1}{c|}{P}&
\multicolumn{1}{l|}{TCP/IP}&
\multicolumn{1}{l|}{Y}& 
\multicolumn{1}{l|}{Y} & 
\multicolumn{1}{l|}{Generic}\\ \hline

\multicolumn{1}{|l}{SSAM~\cite{lee2018application}}&
\multicolumn{1}{c|}{2018} &
\multicolumn{1}{c|}{MM} &
\multicolumn{1}{c|}{C3D} & 
\multicolumn{1}{l|}{S}  & 
\multicolumn{1}{c|}{CPU}&
\multicolumn{1}{l|}{P} &
\multicolumn{1}{l|}{K} & 
\multicolumn{1}{l|}{CPU} &
\multicolumn{1}{c|}{P}&
\multicolumn{1}{l|}{API}&
\multicolumn{1}{l|}{N}& 
\multicolumn{1}{l|}{N} & 
\multicolumn{1}{l|}{Similarity search}\\ \hline 

\multicolumn{1}{|l}{DNN-PIM~\cite{liu2018processing}}&
\multicolumn{1}{c|}{2018} &
\multicolumn{1}{c|}{MM} & 
\multicolumn{1}{c|}{C3D} & 
\multicolumn{1}{l|}{S}  & 
\multicolumn{1}{c|}{CPU + ACC}& 
\multicolumn{1}{l|}{P+F} & 
\multicolumn{1}{l|}{K} & 
\multicolumn{1}{l|}{CPU} & 
\multicolumn{1}{c|}{P+S}& 
\multicolumn{1}{l|}{OpenCL}& 
\multicolumn{1}{l|}{Y}&  
\multicolumn{1}{l|}{N} & 
\multicolumn{1}{l|}{DNN training}\\ \hline 

\multicolumn{1}{|l}{Boroumand~\cite{Boroumand:2018:GWC:3296957.3173177}}& 
\multicolumn{1}{c|}{2018} & 
\multicolumn{1}{c|}{MM} & 
\multicolumn{1}{c|}{C3D} &  
\multicolumn{1}{l|}{S}  & 
\multicolumn{1}{c|}{CPU+ACC}& 
\multicolumn{1}{l|}{P+F} & 
\multicolumn{1}{l|}{K} & 
\multicolumn{1}{l|}{CPU} & 
\multicolumn{1}{c|}{S}& 
\multicolumn{1}{l|}{-}& 
\multicolumn{1}{l|}{Y}&  
\multicolumn{1}{l|}{-}&  
\multicolumn{1}{l|}{Google workloads}\\ \hline

\multicolumn{1}{|l}{CompStor~\cite{torabzadehkashi2018compstor}}& 
\multicolumn{1}{c|}{2018} & 
\multicolumn{1}{c|}{SCM} & 
\multicolumn{1}{c|}{SSD} &  
\multicolumn{1}{l|}{US}  & 
\multicolumn{1}{c|}{CPU}& 
\multicolumn{1}{l|}{P} & 
\multicolumn{1}{l|}{A} & 
\multicolumn{1}{l|}{CPU} & 
\multicolumn{1}{c|}{P}& 
\multicolumn{1}{l|}{API}&
\multicolumn{1}{l|}{-}&  
\multicolumn{1}{l|}{Y} &  
\multicolumn{1}{l|}{Text search}\\ \hline

\end{tabular}

}

\caption{Architectures classification and evaluation, refer to table~\ref{tab:arcLegend} for a legend}
\label{tab:arcClassification}
\end{table*}

\section{Classification and Evaluation}
\label{sec:classificationmodel}
This section introduces the classification and evaluation metrics that are used
in Section~\ref{sec:procmainmem} and~\ref{sec:procstorage} and is summarized in Table~\ref{tab:arcLegend} and Table~\ref{tab:arcClassification}. For each architecture, five main categories are evaluated and classified:
\begin{itemize}
\item{\textit{Memory}} - The decision of using what kind of memory is one of the most fundamental questions on which the near-memory architecture depends.
\item{\textit{Processing}} - Processing unit implemented, and the granularity of processing it performs plays a critical role. 
\item{\textit{Tool}} - Any system's success depends heavily on the available tool support. The availability of tool infrastructure indicates the maturity of the architecture. 
\item{\textit{Interoperability}} - Interoperability deals with programming model, cache coherence, virtual memory, and efficient data mapping. Interoperability is one of the key enablers for the adoption of any new system.
\item{\textit{Application}} - NMC is data-centric and is usually specialized for particular workload. Therefore, in our evaluation, we include the domain of the application.  
\end{itemize}

\section{Processing near Main Memory}
\label{sec:procmainmem}
Processing near main memory can be coupled with different processing units ranging from programmable to fixed-functional units. We describe some of the notable architectures in this class. All solutions discussed in this section are summarized in Table~\ref{tab:arcClassification}.

\subsection{Programmable Unit}
\label{subsec:programmableunitmm}


\textbf{NDC (2014)} Pugsley\etal~\cite{6844483} focus on Map-Reduce workloads, characterized by localized memory accesses and embarrassing parallelism. The architecture consists of a central multi-core processor connected in a daisy-chain configuration with multiple 3D-stacked memories. Each memory houses many ARM cores that can perform efficient memory operations without hitting the memory wall. However, they were not able to fully exploit the high internal bandwidth provided by HMC. Therefore, NMC processing units need careful redesigning to saturate the available bandwidth.

\textbf{TOP-PIM (2014)} Zhang\etal\cite{zhang2014top} propose an architecture based on accelerated processing unit (APU). Each APU consists of a GPU and a CPU on the same silicon die. The APUs are interconnected with high-speed serial links with multiple 3D-stacked memory modules. APU allows code portability and easier programmability. The kernels analyzed span from graph processing to fluid and structural dynamics. The author uses traditional coherence mechanism based on restricted memory regions which puts restriction on data placement. 

\textbf{AMC (2015)} Nair\etal\cite{nair2015active} develop active memory cube (AMC), which is built upon the HMC. They add several processing elements to the vault of HMC and refer to it as``lanes''. Each lane has its register file, a load/store unit performing read and write operations to the memory contained in the same AMC, and a computational unit. The communication between AMCs is coordinated by the host processor. Compiler support based on OpenMP for C/C++ and FORTRAN is provided.

\textbf{PIM-enabled (2015)} Ahn et al.~\cite{ahn2015pim} leverage existing programming model so that the conventional architectures can exploit the PIM concept without changing the programming interface. They implement it by adding compute-capable commands and specialized instruction to trigger the NMC computation. NMC processing units are composed of computation logic (e.g. adders) and an SRAM operand buffer and are housed in the logic layer of the HMC. Offloading at instruction level, however, could lead significant overhead. In addition, the proposed solution requires significant changes on the application side, hence reducing application readiness and hurdle wide adoption.

\textbf{TESSERACT (2015)} Ahn et al.~\cite{7284059} focus on graph processing applications. Their architecture consists of one host processor and an HMC with multiple vaults, which has an out-of-order processor mapped to each vault. These cores can see only their local data partition, but they can communicate with each other using a message passing protocol. The host processor has access to the entire address space of the HMC. To exploit the high available memory bandwidth in the systems, they develop prefetching mechanisms. 

\textbf{TOM (2016)} Hsieh et al.~\cite{7551394} propose an NMC architecture consisting of a host GPU interconnected to multiple 3D-stacked memories that has small light weight GPU cores. They develop a compiler framework that automatically identifies possible offloading candidates. Code blocks are marked as beneficial to be offloaded by the compiler if the saving in memory bandwidth during the offloading execution is higher than the cost to initiate and complete the offload.  A runtime system takes the final decision to where to execute a block. Furthermore, the framework uses a mapping scheme that ensures data and code co-location. The cost function proposed for code offloading makes use of static analysis to estimate the bandwidth saving. Static analysis, however, may fail on code with indirect memory accesses.

\textbf{Pattnaik\footnote{Architecture has no name, first author's name is shown.\label{noAuthor}} (2016) }Pattnaik et al.~\cite{7756764} similar to~\cite{7551394} develop an NMC-assisted GPU architecture. An affinity prediction model decides where a given kernel should be executed while a scheduling mechanism tries to minimize the application execution time. The scheduling mechanism can overrule the decision made by the affinity prediction model. To keep memory consistency between the main GPU and the near-memory cores, the author propose to invalidate the L2 cache of main GPU after each kernel execution. Despite having a simple implementation, it is not optimal as refilling the cache can lead to a considerable overhead.

\textbf{MONDRIAN (2017)} Drumond et al.~\cite{de2017mondrian} show that hardware and software co-design is needed to achieve efficiency and performance for NMC systems. In particular, they show that the current optimization of data-analytic algorithms heavily rely on random memory accesses while NMC system prefer sequential memory accesses to saturate the huge bandwidth available. Based on this observation, the system proposed consists of a mesh of HMC with tightly connected ARM cores in the logic layer.


\textbf{MCN (2018)} Alian et al.~\cite{alian2018application} use a lightweight processor similar to Qualcomm Snapdragon 835 as a near-memory processing unit in the buffered DRAM DIMM. The memory channel network (MCN) processor runs a lightweight OS with network software layers essential for running a distributed computing framework. The most striking feature of MCN is that the authors demonstrate unified near-data processing across various nodes using ConTutto FPGA with IBM POWER8. Supporting the entire TCP/IP stack on the near-memory accelerator requires a complex accelerator design. Current trends in the industry, however, are pushing for a simplified accelerator design shifting the complexity on the cores side (i.e., OpenCAPI~\cite{stuecheli2018ibm}).

\textbf{SSAM (2018)} Lee et al.~\cite{lee2018application} use Pin tool to characterize the architectural behaviors of kNN, which are at the core of similarity search applications. The instruction mix profile reveals a higher percentage of vector operations and memory reads, which confirms that vector operations are important for kNN workloads and are bound by high data movement. Based on the observation, they integrate specialized vector processing units in the logic layer of HMC and propose instruction extensions to leverage those hardware units. Similar to~\cite{ahn2015pim} they require to modify the application and the instruction-set architecture to exploit near-memory acceleration.

\textbf{DNN-PIM (2018)} Liu et al.~\cite{liu2018processing} propose heterogeneous NMC architecture for training of deep neural network models. The logic layer of 3D-stacked memory comprises programmable ARM cores and large fixed-function units (adders and multipliers). They extend the OpenCL programming model to accommodate the NMC heterogeneity. Both fixed-function NMC and programmable NMC appear as distinct compute devices and a runtime system dynamically maps and schedules NN kernels on heterogeneous NMC system, based on online profiling of NN kernels.

\textbf{Boroumand\textsuperscript{\ref{noAuthor}} (2018)} Boroumand \etal\cite{Boroumand:2018:GWC:3296957.3173177} evaluate NMC architectures for Google workloads. They observe that many Google workloads spend a considerable amount of energy in data movement. Based on their observation, they propose two NMC architectures, one based on CPUs and the other one using fixed-function accelerator on top of HBM
While accelerating these Google workloads, they take into account the low area and power budget in consumer devices. They evaluate the benefits of the proposed NMC architectures by extending the gem5 simulator. Differently from previous works they focused on daily applications.


\subsection{Fixed-Function Unit}
\label{subsec:ffunit}

\textbf{JAFAR (2015) }Xi\etal\cite{xi2015beyond} embed an accelerator in a DRAM module to implement database select operations. The key idea is to use the near-memory accelerator to scan and filter data directly in the memory, thus having a significant reduction in data movement. Only the relevant data will be pushed up to the host CPU. To control the accelerator, the authors suggest using memory-mapped registers to read and write via application program interface (API) function calls. Even though JAFAR shows promising potential in database applications, its evaluation is quite limited and it can handle only filtering operations. More complex operations fundamental for the database domain such as sorting, indexing and compression are not considered. 

\textbf{IMPICA (2016)} Hsieh et al.~\cite{hsieh2016accelerating} accelerate pointer chasing operations which are ubiquitous in data structures. They propose adding specialized units that decouple addresses generation from memory accesses in the logic layer of 3D-stacked memory. These units traverse through the linked data structure in memory and return only the final node found to the host CPU. They also propose to completely decouple the page table of IMPICA from the host CPU to avoid virtual memory related issues.  
The memory coherence is assured by demarking different memory zones for the accelerators and the host CPU. This design was able to fully exploit the high level of parallelism in pointer chasing. 

\textbf{Vermij\textsuperscript{\ref{noAuthor}} (2017)} Vermij et al.~\cite{vermij2017sorting} propose a system for sorting algorithm where phases having high temporal locality are executed on the host CPU, while algorithm phases with poor temporal locality are executed on an NMC device. The architecture proposed consists of a memory-technology agnostic controller located at the host CPU side and a memory-specific controller tightly coupled with the NMC system. The NMC accelerators are placed in the memory-specific controllers and are assisted with an NMC manager. The NMC manager also provides support for cache coherency, virtual memory management and communications with the host processor.

\textbf{GraphPIM (2017)} Nai et al.~\cite{nai2017graphpim} map graph workloads in the HMC exploiting its atomic\footnote{Atomic instruction such as compare-and-set are indivisible instructions. The cpu is not interrupted when performing such operations.} functionality. As they focus on atomics, they can offload at instruction granularity. Notably, they do not introduce new instructions for NMC and make use of the host instruction set to map to NMC atomics through an uncacheable memory region. Similar to~\cite{ahn2015pim}, offloading at instruction granularity can have significant overhead. Besides, the mapping to NMC atomics instruction requires the graph framework to allocate data on particular memory regions via custom \textit{malloc}. This requires changes on the application side, reducing the application readiness.

\subsection{Reconfigurable Unit}
\label{subse:reconfigunit}
\textbf{Gokhale\textsuperscript{\ref{noAuthor}} (2015)} Gokhale et al.~\cite{gokhale2015near} propose to place a data rearrangement engine (DRE) in the logic layer of the HMC to accelerate data accesses while still performing the computation on the host CPU. The authors target cache unfriendly applications with high memory latency due to irregular access patterns, e.g., sparse matrix multiplication.  Each of the DRE engines consists of a scratchpad, a simple controller processor, and a data mover. In order to make use of the engines, the authors develop an API with different operations. Each operation is issued by the main application running on the host and served by a control program loaded by the OS on each DRE engine. Similar to~\cite{7756764} the authors propose to invalidate the CPU caches after each fill and drain operation to keep memory consistency between the near-memory processors and the main CPU. As pointed out earlier, this approach can introduce a significant overhead. Furthermore, the synchronization mechanism between the CPU and the near-memory processors is based on polling. This means that the CPU wastes clock cycles waiting for the near-memory accelerator to complete its operations. On the other hand, a lightweight synchronization mechanisms based on interrupts could be used as more efficient alternative.


\textbf{HRL (2015)} Gao et al.~\cite{7446059} propose a reconfigurable logic architecture called heterogeneous reconfigurable logic (HRL) that consists of three main blocks: fine-grained configurable logic blocks (CLBs) for control unit, coarse-grained functional units (FUs) for basic arithmetic and logic operations, and output multiplexer blocks (OMBs) for branch support. Each memory module follows HMC like technology and houses multiple HRL devices in the logic layer. The central host processor is responsible for data partition and synchronization between NMC units. As in the case of~\cite{zhang2014top} to avoid consistency issues and virtual-to-physical translation the authors propose memory-mapped non-cachable memory region putting restriction on data placement.

\textbf{NDA (2015)} Farmahini et al.~\cite{7056040} propose three different NMC architectures using coarse-grained reconfigurable arrays (CGRA) on commodity DRAM modules. This proposal requires minimal change to the DRAM architecture. However, programmers should identify which code would run close to memory. This leads to increased programmer effort for demarking compute-intensive code for execution. Also, it does not support direct communication between NMC stacks.


\section{Processing near Storage Class Memory}
\label{sec:procstorage}
Flash and other emerging nonvolatile memories such as phase-change memory (PCM)~\cite{Qureshi:2009:SHP:1555815.1555760}, spin-transfer torque RAM (STT-RAM)~\cite{1609379}, etc., are termed as storage-class memories (SCM)~\cite{7151782}. These memories are trying to fill the latency gap between DRAMs and disks. SCM like NVRAM is even touted as a future replacement for DRAM~\cite{ranganathan2011microprocessors}. Moving computation in SCM has some of the similar benefits to DRAM concerning savings in bandwidth, power, latency, and energy but also because of the higher density it allows to work on much larger data-sets as compared to DRAM~\cite{quero2015self}.

\subsection{Programmable Unit}
\label{subsec:programmableunitscm}

\textbf{XSD (2013)} Cho et al.~\cite{cho2013xsd} propose a SSD architecture that integrates graphics processing unit (GPU) close to the memory. They provide an API based on the MapReduce framework that allows users to express parallelism in their application, and that exploit the parallelism provided by the embedded GPU. They develop a performance model to tune the SSD design. The experimental results show that the proposed XSD is approximately $25\times$ faster compared to an SSD model incorporating a high-performance embedded CPU.However, the host CPU ISA needs to be modified to launch the computation on the GPU embedded inside the SSD.

\textbf{WILLOW (2014)} Seshadri et al.~\cite{seshadri2014willow} propose a system that has programmable processing units referred to as storage processor units (SPU). Each SPU runs a small operating system that maintains and enforces security. On the host-side, the Willow driver creates and manages a set of objects that allow the OS and applications to communicate with SPUs. The programmable functionality is provided in the form of SSD Apps.  Willow enables programmers to augment and extend the semantics of an SSD with application-specific features without compromising file system protection. The  programming model based on RPC supports the concurrent execution of multiple SSD Apps and the execution of trusted code. However, it neither supports dynamic memory allocation and nor  allow users to dynamically load their tasks to run on the SSD.

\textbf{SUMMARIZER (2017)} Koo et al.~\cite{koo2017summarizer} design APIs that can be used by the host application to offload filtering task to the inherent ARM-based cores inside an SSD processor. This approach reduces the amount of data transferred to the host and allows the host processor to work on the filtered result. They evaluate static and dynamic strategies for dividing the work between the host and SSD processor. However, sharing the SSD controller processor for user applications and SSD firmware can lead to performance degradation due to interference between I/O tasks and in-storage compute tasks.

\textbf{CompStor (2018) } Torabzadehkashi \etal~\cite{torabzadehkashi2018compstor} propose an architecture that consists of NVMe over PCIe SSD and FPGA based SSD controller coupled with in-storage processing subsystem (ISPS) based on the quad-core ARM A53 processor.  They modify the SSD controller hardware and software to provide high bandwidth and low latency data path between ISPS and the flash media interface. Fully isolated control and data paths ensure concurrent data processing and storage functionality without degradation in performance of either one. The architecture supports porting a Linux operating system. However, homogeneous processing cores in the SSD are not sufficient to meet the requirement of complex modern applications~\cite{torabzadehkashi2019catalina}.

\subsection{Fixed-Function Unit}
\label{subsec:ffunit2}

\textbf{Smart SSD (2013)} Kang et al.~\cite{kang2013enabling} propose a model to harness the processing power of the SSD using an object-based communication protocol. They implement the Smart SSD features in the firmware of a Samsung SSD and modify the Hadoop core and MapReduce framework to use task-lets as a map or a reduce function. To evaluate the prototype, they used a micro-benchmark and log analysis application on both a device and a host. Their SmartSSD was able to outperform host-side processing drastically by utilizing internal application parallelism. Likewise, Do~et~al.~\cite{park2014query} extend Microsoft SQL Server to offload database operations onto a Samsung Smart SSD. The selection and aggregation operators are compiled into the firmware of the SSD. Quero~et~al.~\cite{quero2015self} modify the Flash Translation Layer (FTL) abstraction and implement indexing algorithm based on the B++tree data structure to support sorting directly in the SSD. The approach of modifying the SSD firmware to support processing of certain functions is fairly limited and can not support the wide variety of workloads~\cite{torabzadehkashi2019catalina}.

\textbf{ProPRAM (2015)} Wang et al.~\cite{wang2015propram} observe that NVM is often naturally incorporated with basic logic like data comparison write or flip-n-write module, and exploit the existing resources inside memory chips to accelerate the critical non-compute intensive functions of emerging big data applications. They expose the peripheral logic to the application stack through ISA extension. Similar to~\cite{kang2013enabling,park2014query,quero2015self}, this approach cannot support the diverse workloads requirements.

\textbf{BISCUIT (2016)} Gu et al.~\cite{gu2016biscuit} present a near-memory computing framework, that allows programmers to write a data-intensive application to run in a distributed manner on the host system and the storage system. The SSD hardware incorporates a pattern matcher IP designed for NMC. The NMC system, however, acts as a slave to the host CPU and all data is controlled by the host CPU. When this hardware IP is applied, modified MySQL significantly improves TPC-H performance. Similar to~\cite{koo2017summarizer} two real-time ARM cores in the SSD are shared between the user runtime and SSD firmware.

\subsection{Re-configurable Unit}
\label{subse:reconfigunit2}

\textbf{BlueDBM (2015) }Jun et al.~\cite{Jun:2015:BAB:2872887.2750412} present a flash-based platform, called BlueDBM, built of flash storage devices augmented with an application specific FPGA based in-storage processor. The data-sets are stored in the flash array and are read by the FPGA accelerators. Each accelerator implements a variety of application-specific distance comparators, used in the high-dimensional nearest-neighbor search algorithms. They also use the same architecture exploration platform for graph analytics. Authors propose sort-reduce algorithm to solve the random read-modify update into vertex data in the vertex-centric programming model. The algorithm is then accelerated on FPGA that uses flash memory for edges, vertices and partially sort-reduced files~\cite{jun2018grafboost}. However, the platform does not support dynamic task loading, similar to ~\cite{seshadri2014willow}, and has limited OS-level flexibility. 

\textbf{CARIBOU (2017) }Zsolt et al.~\cite{istvan2017caribou} enable key-value store interface over TCP/IP socket to the storage node comprising of FPGA connected to DRAM/NVRAM. They implement selection operators in the FPGA, which are parameterizable at runtime, both for structured and unstructured data to reduce data movement and to avoid the negative impact of near-data processing on the data retrieval rate. Similar to~\cite{Jun:2015:BAB:2872887.2750412} Caribou does not have OS-level flexibility, e.g. file-system is not supported transparently

\subsection*{Discussion}
From table 2, one can see that the majority of research papers published over the years have proposed homogeneous processing units near the memory. The logic considered varies in type e.g. simple in-order cores~\cite{torabzadehkashi2018compstor,lee2018application,alian2018application,de2017mondrian,koo2017summarizer,7927081,7284059,nair2015active,6844483}, graphics processing units~\cite{zhang2014top,7756764,7551394,cho2013xsd}, field programmable gate arrays~\cite{7446059, Jun:2015:BAB:2872887.2750412,istvan2017caribou} and application specific accelerators~\cite{xi2015beyond,gokhale2015near,ahn2015pim,hsieh2016accelerating,gu2016biscuit,vermij2017sorting,nai2017graphpim,liu2018processing}. Majority of the NMC proposals are targeted towards different types of data processing applications e.g. graph processing~\cite{zhang2014top,7284059,nai2017graphpim}, MapReduce~\cite{cho2013xsd,6844483,7056040}, machine learning~\cite{liu2018processing,lee2018application}, database search~\cite{kang2013enabling,xi2015beyond,gu2016biscuit,istvan2017caribou,7927081,koo2017summarizer}. The logic layer connected to the memory via silicon vias in 3D stacked memories is considered to be the most important play ground of innovation ~\cite{Boroumand:2018:GWC:3296957.3173177,liu2018processing,lee2018application,de2017mondrian,vermij2017sorting,7756764,7551394,hsieh2016accelerating,ahn2015pim,7446059,gokhale2015near,7284059,nair2015active,zhang2014top,6844483} but the efficacy of those new innovations has only been tested using simulators. On the other hand, the integration of NMC units near the traditional DRAM and SSD controllers have been emulated using FPGAs~\cite{gu2016biscuit, awan2017performance, koo2017summarizer, istvan2017caribou,kang2013enabling,seshadri2014willow}. Despite the promises made by existing proposals on NMC, the support for memory consistency and virtual memory is fairly limited and the programmers are expected to re-write their code using specialized APIs in order to reap the benefits of NMC~\cite{torabzadehkashi2018compstor, lee2018application, nai2017graphpim, de2017mondrian, koo2017summarizer,vermij2017sorting,istvan2017caribou,gu2016biscuit}. 

The idea of populating homogeneous processing units near the memory to accelerate a specific class of workloads is limited in a sense that NMC enabled servers deployed in the data centers are expected to host a wide variety of workloads. Hence, these systems would need heterogeneous processing units near the memory~\cite{Boroumand:2018:GWC:3296957.3173177,liu2018processing,7446059} to support the complex mix of data center workloads. For broader adoption of the NMC by the application programmers, methods that enable transparent offloading to the NMC units would also be needed. Transparent offloading requires the compiler or the run-time system to identify code regions based on some application characteristics. One of the most used metrics is the number of last-level cache misses~\cite{Hadidi:2017:CCT:3154814.3155287, awan2017identifying,awan2017performance, gokhale2015near, ahn2015pim}. Unfortunately, the integration of a profiler (such as Perf or Pin) in a compiler or run-time system is still a challenging task~\cite{Hadidi:2017:CCT:3154814.3155287} due to its dynamic nature. Therefore, current solutions rely on commercial profiling tools, such as Intel VTune~\cite{Vtune}, to detect the offloading kernels~\cite{awan2017performance, awan2017identifying, hsieh2016accelerating}. Additionally, some works~\cite{zhang2014top, liu2018processing, hsieh2016accelerating} use the bandwidth saving as an offloading metric. 

\section{Challenges of Near-Memory Computing}

\label{sec:interondpcachevirtual}
In this section we explain the challenges related to programming model, data mapping, virtual memory support, and cache coherency that NMC needs to address before it can be established as a de facto solution for HPC. Additional challenges on design space exploration, reliable simulators, which can deal with these heterogeneous environments, will be discussed in Section~\ref{sec:modelingmethod}.

\subsection{Virtual Memory Support}
\label{subsec:virtualmemndp}
Support for virtual memory is achieved using: \textit{paging} or \textit{direct segmentation}. 

\textit{1) Paging} can be software or hardware managed. Most NMC systems adopt a software-managed translation lookaside buffer (TLB) to provide a mapping between virtual and physical addresses~\cite{7429299,Sura:2015:DAO:2742854.2742863,7551394,Wei05anear-memory}. Other works such as~\cite{Azarkhish:2016:DEP:2963802.2963805} observe that an OS managed TLB may not be the optimal solution and propose a hardware implementation where a simple controller is responsible for fetching the entry in the page table. In~\cite{hsieh2016accelerating} Hsieh~et~al. notice that for pointer-chasing applications, the accelerator works only on specific data structures that can be mapped onto a contiguous region of the virtual address space. As a consequence, the virtual-to-physical translation can be designed to be more compact and efficient. Picorel\etal\cite{picorel2017near} show that restricting the associativity such that a virtual page can be mapped only with few contiguous physical pages can break the decode-and-fetch serialization. Their translation mechanism, DIPTA, allows removing the virtual-to-physical translation overhead.

\label{subsubsec:directsegment} 
 \textit{2) Direct Segmentation}~\cite{7459501,7459537} consists in a simplified approach where part of the linear virtual address is mapped to physical memory using a direct segment rather than a page. This mechanism allows to remove TLB misses overhead and greatly simplifies the hardware~\cite{Basu:2013:EVM:2508148.2485943}.

\subsection{Cache Coherence}
\label{subsec:cachecoherencendp}
Cache coherence is one of the most critical challenge in adoption of near-memory computing. Depending upon the coherence mechanism used, drastically changes on performance and programming model can be obtained. The two primary mechanisms proposed in the literature to achieve cache coherency support are \textit{restricted memory region} and \textit{non-restricted region}.

\textit{1) Restricted region} techniques such as the on used by Farmahini~\etal~\cite{7056040} divides the memory into two parts: one for the host processor and one for the accelerator, which is uncacheable. Ahn\etal~\cite{7284059} has used a similar approach for graph processing algorithms. 
Another strategy proposed by Ahn\etal\cite{ahn2015pim} provides a simple hardware-based solution in which the NMC operations are restricted to only one last level cache block, due to which they can monitor the cache block and request for invalidation or write-back if required.

\textit{2) Non-Restricted Memory Regions}
 on the other hand can potentially lead to a significant amount of memory traffic. Pattnaik\etal\cite{7756764} propose to maintain coherence between the main/host GPU and near-memory compute units, flushing the L2 cache in the main GPU after kernel execution. With this approach, potentially useful data could be evicted from cache. Another way is to implement a look-up based approach as done by Hsieh\etal\cite{7551394}. The NMC units record the cache line address that has been updated by the offloaded block, and once the offloaded block is processed, they send this address back to the host system. Subsequently, the host system gets the latest data from memory by invalidating the reported cache lines. Liu\etal\cite{liu2018processing} propose single global memory with a relaxed consistency model, shared between NMCs and CPU. They implement explicit synchronization points to synchronize the accesses to shared variables across NMCs and CPU. More recently, Boroumand\etal\cite{Boroumand:2019:CEC:3307650.3322266} analyze current state-of-the-art coherence mechanisms and show that for NMC a majority of the off-chip traffic due to coherency is unnecessary and can be eliminated if the cache coherency mechanism itself has insight on the accelerator accesses. Based on such observation they propose CoNDA.

\subsection{Programming Model}
\label{subsec:programmingmodel}

A critical challenge in the adoption of NMC is to support a heterogeneous processing environment of a host system and near-memory processing units. It is not trivial to determine which part of an application should run on the near-memory processing units. Work such as~\cite{Hadidi:2017:CCT:3154814.3155287, 7551394} leave this effort on the compiler, while \cite{ghose2018enabling, xi2015beyond, ahn2015pim} on the programmer. Another approach~\cite{ahn2015pim, lee2018application, nai2017graphpim} uses some special set of NMC instructions which invokes NMC logic units. This approach, however, calls for a sophisticated mechanism as it touches most of the software stack from the application down to instruction selection in the compiler. Run-time systems capable of dynamically profiling applications to identify the potential for NMC acceleration during the first few iterations have also been proposed~\cite{liu2018processing, 7551394}. However, there is still a lot of research required in coming up with an efficient approach to ease the programming burden.

\subsection{Data mapping}
\label{subsec:datamappNDPaPIM}
The absence of adequate data mapping mechanisms can severely hamper the benefits of processing close to memory. The data should be mapped in such a way that the data required by the near-memory processing units should be available in the vicinity (data and code co-location). Hence, it is crucial to look into effective data mapping schemes. Hsieh~et~al. in~\cite{7551394} propose a software-hardware co-design method to predict which pages of the memory will be used by the offloaded code, and they tried to minimize the bandwidth consumption by placing those pages in the memory stack closest to the offloaded code. Yitbarek~et~al.~\cite{7459537} propose a data-mapping scheme to place contiguous addresses in the same vault allowing accelerators to access data directly from their own local vault. Xiao et al.~\cite{xiao2018prometheus} propose to model an application with a two-layered graph through the LLVM's Intermediate Representation, distinguishing between memory and computation operations. Once the application's graph is built, their framework, detects groups of vertices called communities, that have a higher probability of connection with each other. Each community is mapped to different vaults.

\begin{figure}[b]
\centering
\includegraphics[width=.95\linewidth]{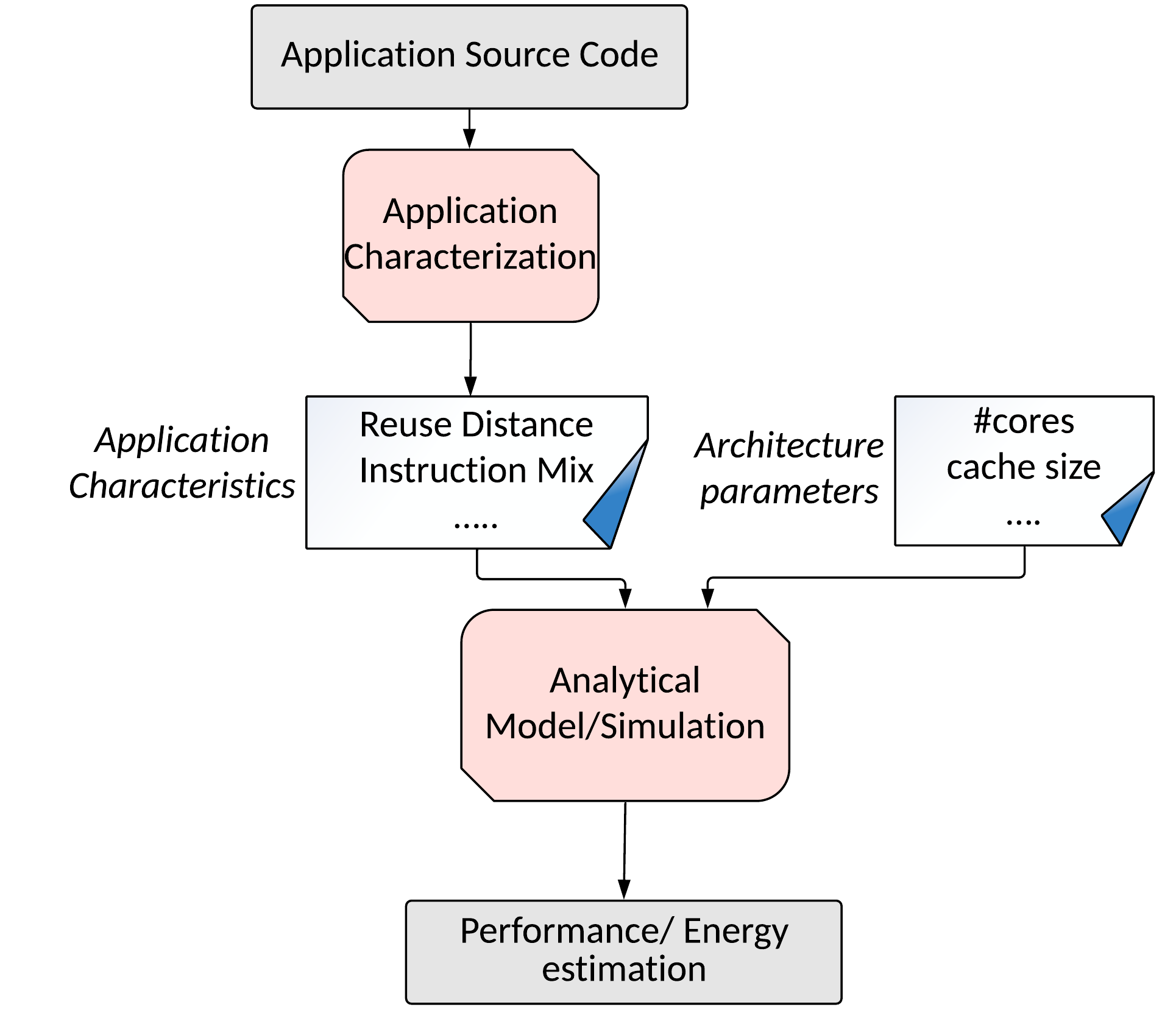}
\caption{Design space exploration highlighting application characteristic with performance evaluation technique
\label{fig:methodology}}
\end{figure}

\section{Design Space Exploration for NMC}
\label{sec:modelingmethod}

As will be clear from the previous classification section, the design space of NMC is huge. To understand and evaluate this space, effective design space exploration (DSE) for NMC systems is required.  Moreover, these architectures calls for specialized hardware software co-design strategies as shown in Figure~\ref{fig:methodology}.

\subsection{Application Characterization}
\label{subse:appchar}

More than ever, application characterization has taken a crucial role in systems design due to the increasing number of new big-data applications. Application characterization is used to extract information by using specific metrics, to decide for a certain application which architecture could have the best performance and energy efficiency. NMC systems have mostly proven to be effective against memory-intensive applications and are hence tailored towards this kind of workloads. To map the hardware to a specific workload, it is critical to analyze the application in order to quantify computational demand and memory footprint. Application characterization can be done either in microarchitecture-dependent or microarchitecture-independent manner.

\textit{1) Microarchitecture-Dependent Characterization} is done using hardware performance counters. Awan et al.~\cite{awan2017identifying} use hardware performance counters to characterize the scale-out big data processing workloads into CPU-bound, memory-bound, and I/O-bound applications and propose programmable logic near DRAM and NVRAM. However, the use of hardware performance counters is limited by the impact of micro-architecture features like cache size, issue width, etc~\cite{6557175}. To overcome this problem, recently there has been a push towards following an ISA independent characterization approach.

\textit{2) Microarchitecture-Independent Characterization} or ISA independent characterization consists in profiling instruction traces and collect inherent application information such as operation mix, memory behavior in a microarchitecture independent way. This approach provides more relevant workload characteristics than performance counters~\cite{4292057}. 
\subsection{Performance Evaluation Techniques}
\label{subsec:performanceevaluation}
Architects often make use of various evaluation techniques to navigate the design-space, avoiding the cost of chip fabrication. Based on the level of detail required, architects make use of analytic models, or more detailed functional or cycle accurate simulators. As the field of NMC does not have very sophisticated tools and techniques, researchers often spend much time building the appropriate evaluation environment~\cite{junior2017generic,NAPEL}. Additionally, there is a critical need for near-memory specific benchmarks.
 
\textit{1) Analytic models} abstract low-level system details and provide quick performance estimates at the cost of accuracy. In the early design stage, system architects are faced with large design choices which range from semiconductor physics and circuit level to micro-architectural properties and cooling concerns~\cite{jongerius2017analytic}. Thus, during the first stage of design-space exploration, analytic models can provide quick estimates. 

Mirzadeh~et~al.~\cite{mirzadeh2015sort} study a join workload, on multiple HMC like 3D-stacked DRAM devices connected via SerDes links using a first-order analytic model. Zhang~et~al.~\cite{xu2015scaling} design an analytic model using machine learning techniques to estimate the final device performance. Recently, Lima~et~al.~\cite{deLima:2018:DSE:3203217.3203280} provide a theoretical design space study for enabling processing in the logic layer of HMC taking into consideration area and power limits.

\begin{table}[ht]
\renewcommand{\arraystretch}{1.2}
\centering
\begin{tabular}{lccc}
\hline
\textbf{Simulator} & \textbf{Year} & \textbf{Category} & \textbf{NMC capabilities} \\ \hline
Sinuca~\cite{7336224} & 2015 & Cycle-Accurate & Yes \\
HMC-SIM~\cite{7529923} & 2016 & Cycle-Accurate & Limited \\
CasHMC~\cite{7544479} & 2016 & Cycle-Accurate & No \\
SMC~\cite{Azarkhish:2016:DEP:2963802.2963805} & 2016 & Cycle-Accurate & Yes \\
CLAPPS~\cite{junior2017generic} & 2017 & Cycle-Accurate & Yes \\ 
Ramulator-PIM~\cite{ramulator-pim-repo}      & 2019    & Cycle-Accurate & Yes  \\

\hline
\end{tabular}
\caption{Open source simulators}
\label{tab:simulator}
\end{table}

\textit{2) Simulation Based Modeling} allows to achieve more accurate performance numbers. Architects often resort to modeling the entire micro-architecture precisely. This approach, however, can be quite slow compared to analytic techniques. In Table~\ref{tab:simulator} we have mentioned some of the academic efforts to build open-source NMC simulator. Hyeokjun~et~al.~\cite{DBLP:journals/corr/ChoeLPKCY16} evaluate the potential of NMC for machine learning (ML) using a full-fledged simulator of multi-channel SSD that can execute various ML algorithms on data stored on the SSD. Similarly, Jo~et~al.~\cite{jo2016data} develop the iSSD simulator based on the gem5 simulator~\cite{Binkert:2011:GS:2024716.2024718}. Ranganathan~et~al.~\cite{chang2012limits, ranganathan2011microprocessors} for their nano-stores architecture use a bottom-up approach where they build an analytic model which breaks applications down into various phases, based on compute, network, and I/O subsystem activities. The model takes input from the low-level performance and power models regarding the performance of each phase. Then they used COTSon~\cite{cotson-osr09} for detailed micro-architectural simulation.

\section{Practicalities of Near-Memory Computing}
\label{sec:system}
In this section, we describe our approach
to solve some of the issues related to NMC. Besides the design of the near-memory computing platform itself, we are focusing on how such a hybrid system can be effectively programmed to maximize performance and minimize power consumption. Additionally, we are extensively analyzing applications to assess essential application metrics for these devices. The various components of our framework are outlined below.

\subsection{NMC Architecture}
\begin{figure}[h]
\centering
\includegraphics[width=21pc]{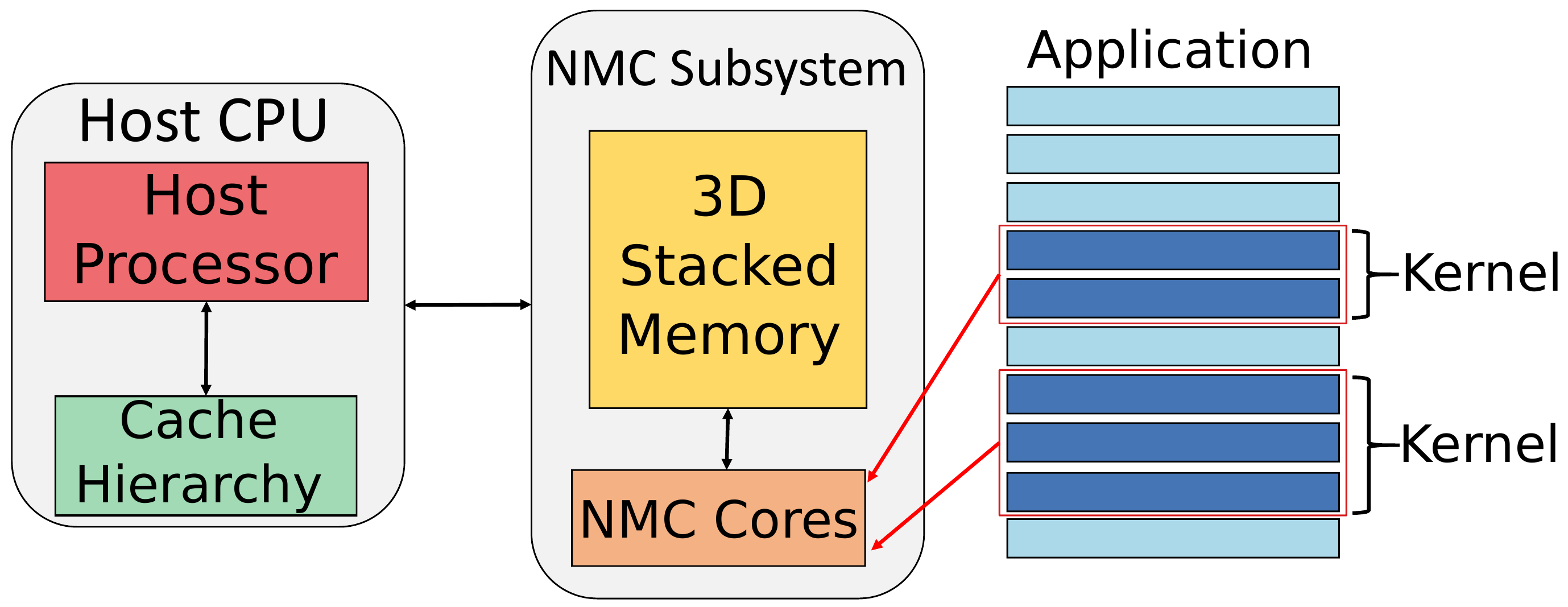}
\caption{Overview of a system with NMC capability~\cite{NAPEL}. On the right, an abstract view of application code with kernels {that are} offloaded to NMC
\label{fig:targetplatform}}
\end{figure}

Figure~\ref{fig:targetplatform} depicts an abstract view of our reference computing platform that we consider. We modeled a multi-core host system interconnected to an external stacked near-memory system via high speed interconnect. In this model, the near-memory compute units are modeled as low power cores (Table~\ref{tab:systemparameters}), which are placed in the logic layer of the 3D stacked DRAM. For the 3D stacked memory, we consider a 4-GB HMC like design, with 8 DRAM layers. Each DRAM layer is split into 16 dual banked partitions with four vertical partitions forming a vault. Each vault has its vault controller.

\begin{table}[h]
\renewcommand{\arraystretch}{1.2}
\centering
\begin{tabular}{l*{3}{c}r}
\hline
\textbf{Description}              & \textbf{Symbol} & \textbf{ARM} \\
\hline
Cores & N\textsubscript{cores} & 4  \\
Host core frequency & F\textsubscript{core} & 3 GHz   \\
NMC core frequency & F\textsubscript{core} & 1.2 GHz   \\
L1 size           & S\textsubscript{l1} & 32 KB  \\
L2 size     & S\textsubscript{l2} & 256 KB  \\
DRAM size & S\textsubscript{DRAM} & 4 GB \\
L1 cache bandwidth & BW\textsubscript{l1} & 137 GB/s\\
L2 cache bandwidth & BW\textsubscript{l2} & 137 GB/s\\
L1 cache hit latency & L\textsubscript{l1} & 1 cycles\\
L2 cache hit latency & L\textsubscript{l2} & 2 cycles\\
\hline
\end{tabular}
\caption{System parameters}
\label{tab:systemparameters}
\end{table}

\subsection{Platform Agnostic Application Characterization for NMC}
\label{subsec:platform-agnostic-NMC}

We define the following microarchitecture independent metrics to characterize the applications from NMC perspective and identify the kernels that can potentially benefit from NMC. 

\begin{itemize}
    \item \textbf{Memory Entropy:} It measures the randomness of the memory accesses. Yen~et~al.~\cite{yen2008notary} devise the formula by applying Shannon's definition to the memory addresses. The higher the memory entropy, the higher the miss ratio. Hence, an application with higher entropy can benefit from NMC.  
    \item \textbf{Spatial Locality:} It measures the probability of accessing nearby memory locations. Spatial Locality is derived from data temporal reuse, which is the number of unique addresses accessed since the last reference of the requested data. We use the formula proposed by Gu~et~al.~\cite{gu2009component} to calculate spatial locality score.     
    \item \textbf{Data-Level Parallelism:} It measures the average possible vector length and is relevant for NMC when employing specific SIMD processing units as near-memory computation units. It is derived from instruction-level parallelism (ILP) score per opcode such as load, store, etc. This metric represents the number of instructions with the same opcode that could run in parallel, thus expressing the DLP per opcode. We compute the average value of DLP using the weighted sum over all opcodes of the DLP per code. 
    \item \textbf{Basic-block level Parallelism:} A basic block is the smallest component in the LLVM intermediate representation (IR) that can be parallelized. If an application has high basic-block level parallelism, it can benefit from multiple processing units.

\end{itemize}

\begin{figure}[] 
    \centering
    \subfloat[Spatial Locality]{%
        \includegraphics[width=8.5cm,trim={2.1cm 2cm 4.4cm 1.8cm},clip]{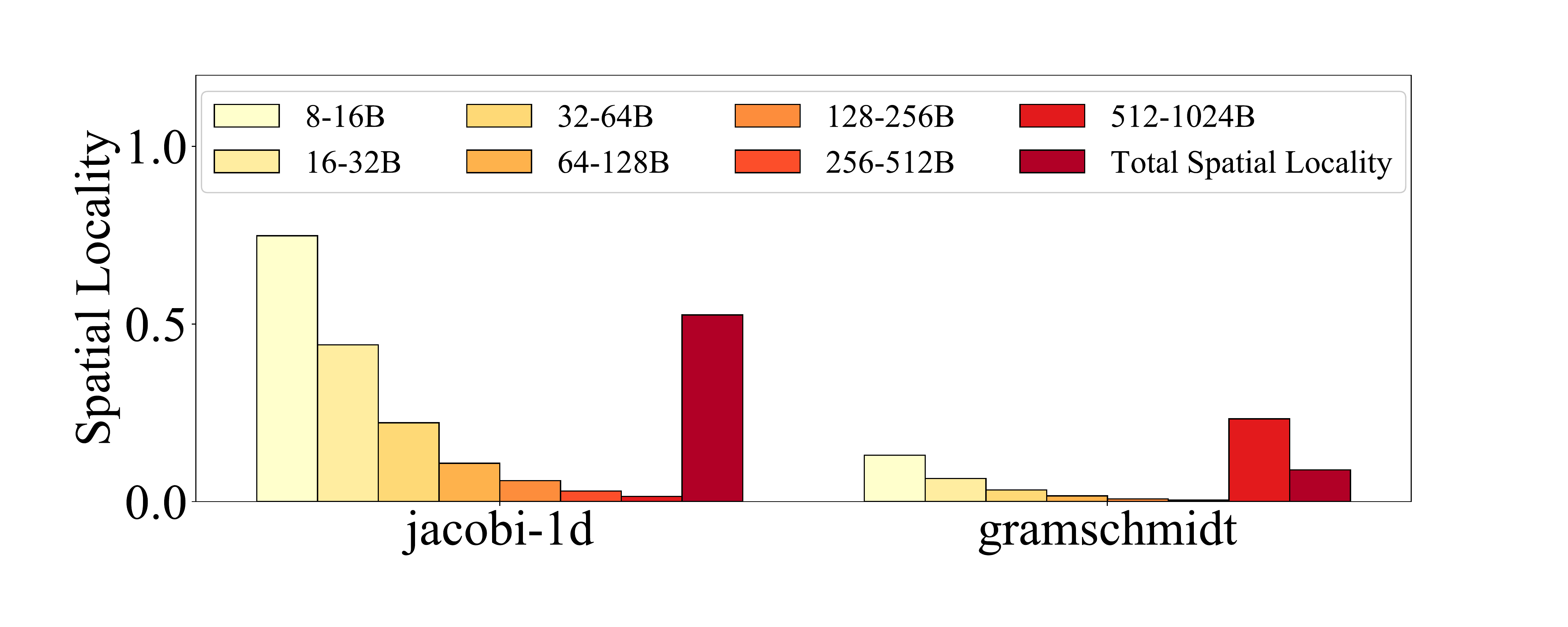}%
        \label{fig:spatial}%
        }%
    \hfill%
    \subfloat[Memory Entropy]{%
        \includegraphics[width=8.5cm,trim={2.1cm 2cm 4.4cm 1.8cm},clip]{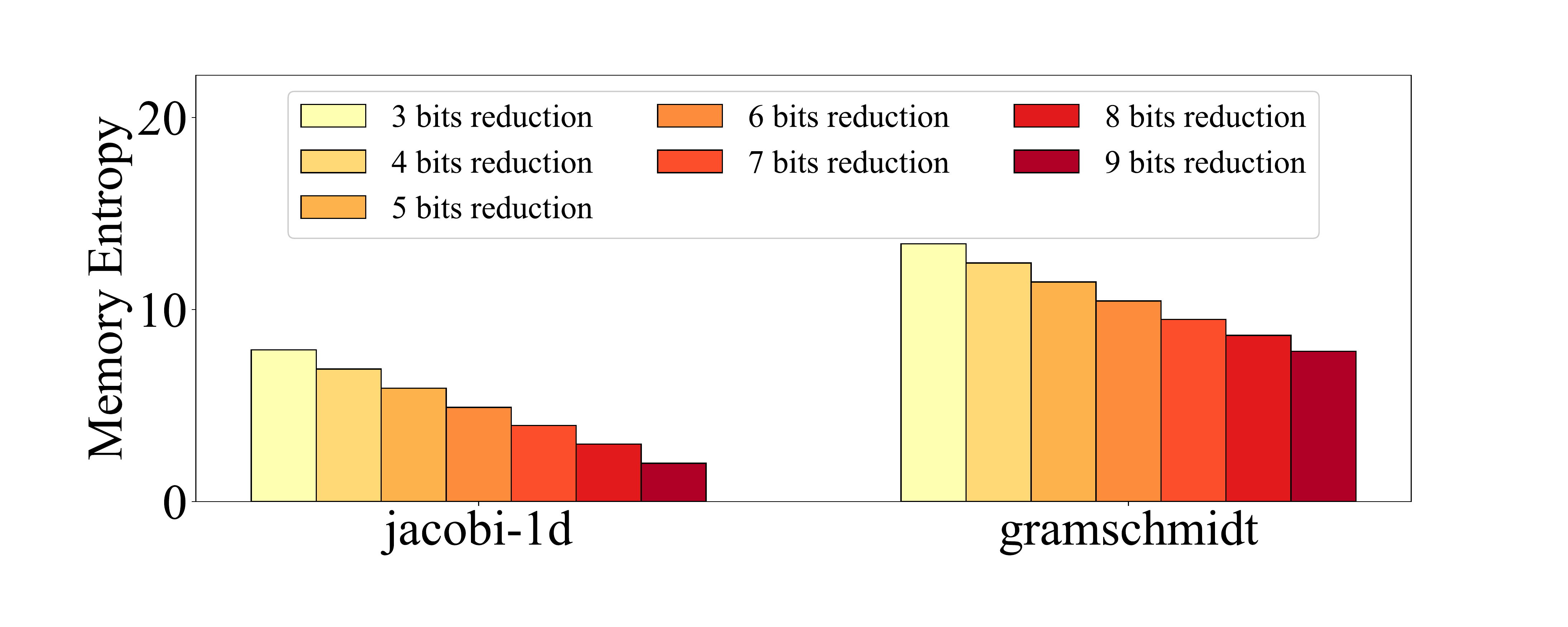}%
        \label{fig:entropy}%
        }%
    \caption{Spatial Locality and Memory Entropy characterization for two applications from PolyBench 
    \label{fig:pisa-nmc}}
\end{figure}

We extend an open-source platform-independent software analysis tool (PISA)~\cite{anghel2016instrumentation} with the above-listed metrics and use it to analyze multi-thread applications from NMC perspective. As an example, we show in Figure~\ref{fig:pisa-nmc} the analysis result of two applications from PolyBench/C 4.1~\cite{pouchet2012polybench}, which is a collection of standard kernels. Each kernel is a single file that can be tuned at compile time. In Figure~\ref{fig:pisa-nmc} we show the characterization of two applications having opposite memory behavior: \texttt{Gramschmidt} and \texttt{Jacobi-1d}.
Figure~\ref{fig:spatial} presents the spatial locality values for different pairs of cache line sizes. For instance, 8-16B means spatial locality doubling the cache line size from 8B to 16B. In the figure, the total spatial locality is the weighted sum of all the values (see \cite{corda2019scopes}). Figure~\ref{fig:spatial} shows, for different cache size configurations, how \texttt{Gramschmidt} has significantly lower spatial locality than \texttt{Jacobi-1d}. This lower locality can be attributed to non-regular memory accesses (e.g., diagonal accesses in a matrix). Figure~\ref{fig:entropy} presents memory entropy for different address granularity. The bit reductions mean reduction in the number of bits of the addresses considered to compute the entropy. The bit reduction reflects as using bigger cache line sizes and could be used to perform further spatial locality analysis (see \cite{corda2019scopes,corda2019DSD}). Figure~\ref{fig:entropy} highlights the lower memory entropy of \texttt{Jacobi-1d} compared to \texttt{Gramschmidt} due to localized memory accesses.
A detailed study on workload characterization from NMC perspective using above mentioned architecture-independent metrics and representative benchmarks are presented in \cite{corda2019scopes,corda2019DSD}.

\begin{figure*}
    \centering
    \includegraphics[width=0.85\linewidth]{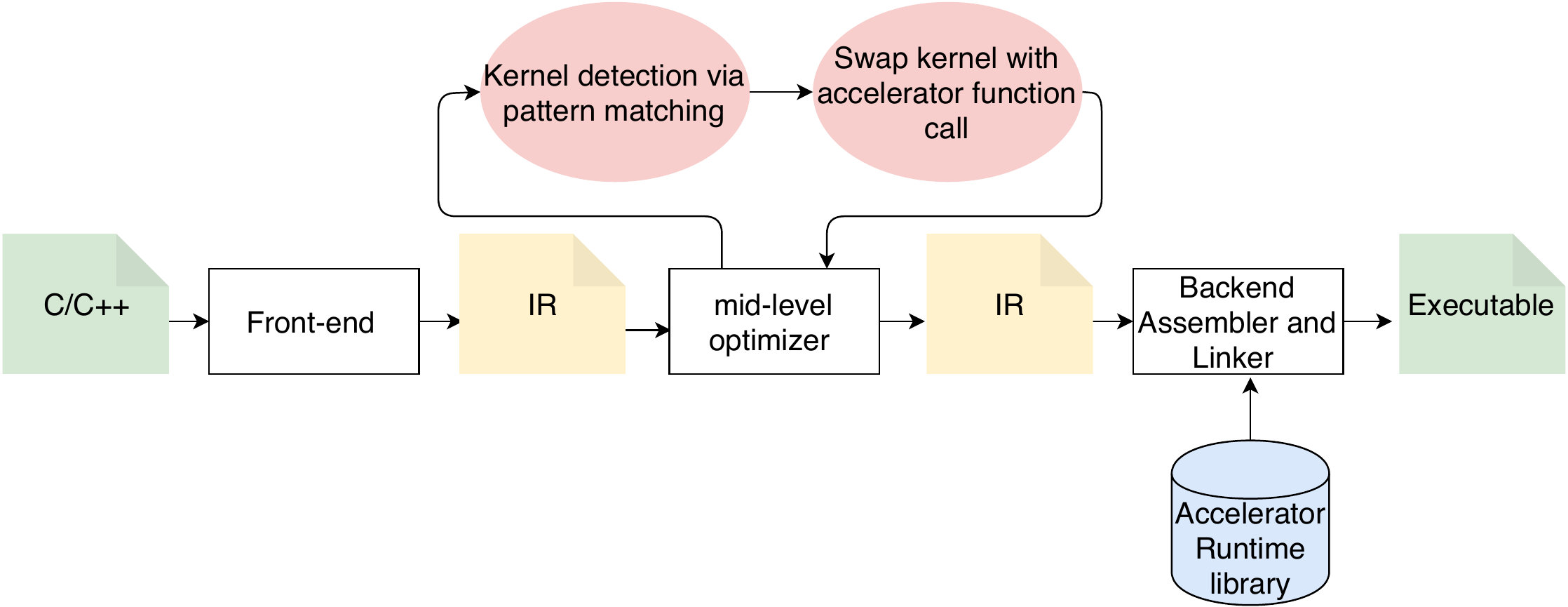}
    \caption{Proposed compilation flow for the system with NMC capabilities depicted in Figure~\ref{fig:targetplatform}. 
    }
    \label{fig:my_label}
\end{figure*}

\subsection{Compilation and Programming Model}
The compiler support has been built on top of our previous work~\cite{zinenko:hal-01965599} and is completely integrated into a state-of-the-art loop optimizer~\cite{grosser2012polly}. Similar to previous works~\cite{akin2015data, 7856642}, we assume our near-memory accelerators expose a library of memory-bound kernels via an application programming interface (API). We provide compiler support to automatically, and transparently for the application,  detect such kernels and call the proper accelerator routine. The compilation flow (Figure~\ref{fig:my_label}) follows a classical compiler design with a front-end, a mid-level optimizer, and target-specific back-ends. We extend this design embedding a state-of-the-art pattern matching framework in the mid-level optimizer. 

Briefly, the compilation flow is as follow: We start from an application written in C/C++; the front-end is responsible for lowering the application code to an intermediate representation. For our work, we use the LLVM compiler infrastructure and exploit its intermediate representation (LLVM-IR). At IR level, we rely on a state-of-the-art polyhedral optimizer~\cite{grosser2012polly} to extract compute kernels and obtain for each of them a mathematical description based on Presburger\cite{10.1007/978-3-642-15582-6_49} sets. On top of this mathematical model, our pattern matching framework is used to match and swap kernels with accelerator-specific run-time calls. Once the run-time calls have been introduced, we rely again on the polyhedral optimizer to lower the mathematical abstraction back to LLVM-IR. During the latest stages of compilation, the run-time accelerator specific libraries are linked with the application binary. With our approach, legacy code can exploit, in a completely transparent way for the user, near-memory acceleration without any change on the application side.

\subsection{Early Design Stage Modeling}
We make use of a first-order analytic model to evaluate the benefit of using near-memory processing in our system in the early design stage. As mentioned in Section~\ref{subsec:performanceevaluation}, a high-level analytic model abstract the detailed micro-architecture. Hence, this approach makes our analytical model more generic and provides quick insights.



Our analytic model incorporates evaluation methodology as described in Figure~\ref{fig:methodology}. To see the potential of computing close to the memory and when it is useful to follow a data-centric approach, we make a comparison of a CPU-centric approach consisting of a multi-core system and compare it to a data-centric approach which has near-memory processing units in the logic layer of HMC like 3D stacked. In the multi-core system, each core has its L1 cache and a shared L2 cache.  In the near-memory system, we assume certain parts of the computation called NMC-kernels would be offloaded to the near-memory compute units, and the other parts will be executed in the host system. The number of vaults inside the HMC and external interconnects links are kept as varying parameters which have an impact on internal and the external bandwidth of the 3D memory respectively.

\begin{figure*}[h]
\centering
\subfloat[Normalized Delay]{\includegraphics[width=9cm]{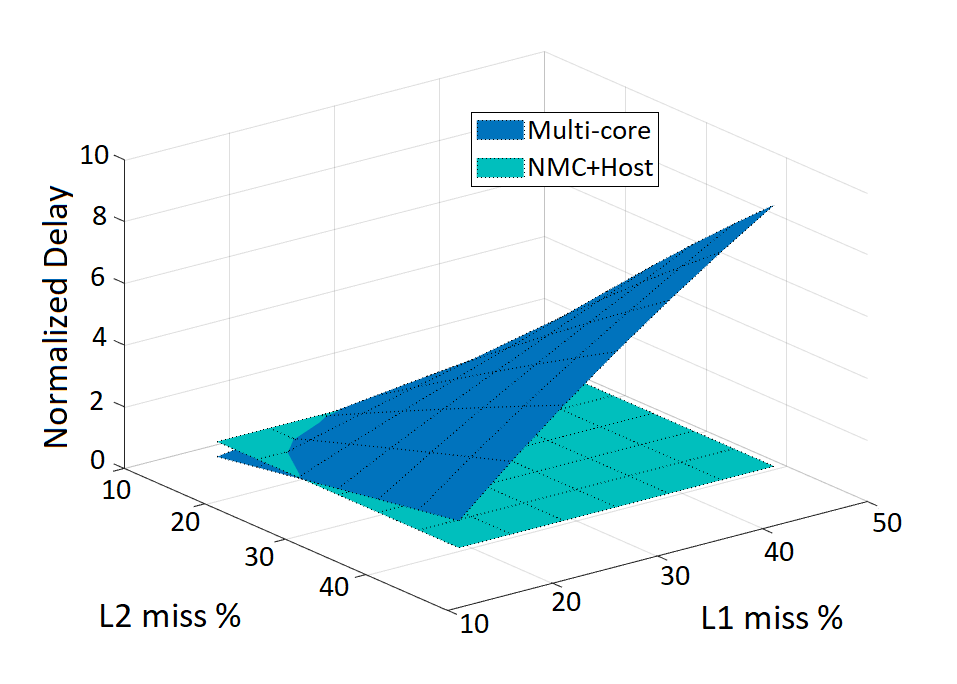}
\label{fig:delay}}
\subfloat[Normalized Energy]{\includegraphics[width=9cm]{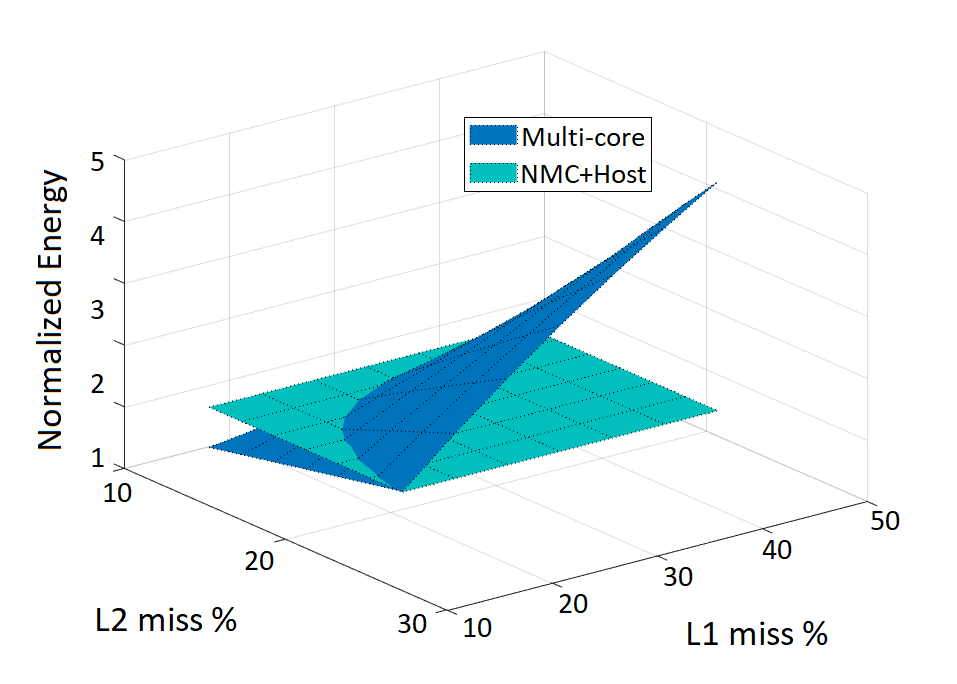}
\label{fig:energy}}
\caption{Performance and energy comparison between multi-core and an NMC system, which is attached to a host system, using high-level analytic model}
\label{fig:results}
\end{figure*}

\textit{Performance Calculation:} The model estimates the execution time as the ratio between the number of memory accesses required to move the block of data to and from the I/O system and the available I/O bandwidth of each memory subsystem. The execution time takes into account the time to process the instructions $\Pi_{non-memory}$ and latency due to memory access $\Pi_{memory}$ (Table~\ref{tab:systemparameters}).


\textit{Energy Calculation:} We build a power model for both CPU-centric and a near-memory system by taking into account dynamic as well as static energy. Dynamic energy is the product of the energy per memory access and the number of memory accesses at each level of the memory hierarchy.  It is assumed to scale linearly with utilization. Static energy is estimated as the product of time and static power for each component. It is scaled based on the number of cores and the cache size. 
    
For modeling the HMC, we refer to~\cite{6242474} and~\cite{6844483}. We consider energy per bit of 3.7 pJ/b for accessing the DRAM layers and 1.5 pJ/b for the operations and data movements in the HMC logic layer. A static power of 0.96 W is assumed~\cite{6844483} to model the additional components in the logic layer.


\textit{Performance and Energy Exploration:} 
We vary the cache miss rate, which encapsulates application characteristics, both at L1 and L2 level to see the impact it has on the performance of the entire system. In case of a near-memory system, we consider the scenario when all accesses are done to the different vaults inside the HMC memory, which can exploit the inherent parallelism offered by the 3D memory. In Figure~\ref{fig:results}, the x and y-axis represent the miss rate at L1 and L2, respectively. Results in Figure~\ref{fig:results} are normalized to NMC+host system. 

Based on our evaluation, we make the following three observations. First, in case of performance and energy, we can see a similar trend, i.e., if the miss rate (L1 and L2) increases the multi-core system performance degrades compared to the NMC systems. This degradation is because of the increase in data movement. The data has to be fetched from the off-chip DRAM. Second, if the CPU does not reuse the data brought into the caches, the use of caching becomes inefficient. Third, the application (or function) with low locality can take advantage of the near-memory approach, whereas the other application (or function) with high locality would benefit more from the traditional approach.


\section{Open Issues and Future Directions}

\label{sec:futuredirections}
Based on our analysis of many existing NMC architectures, we identify the following key
topics for future research that we regard as essential to unlock the full potential of processing close to memory:
\begin{itemize}
   
\item It is unclear which emerging memory technology best supports near-memory architectures; for example, much research is going into new 3D stacked DRAM and non-volatile memories such as PCM, ReRAM, and MRAM. The future of these new technologies relies heavily on advancements in endurance, reliability, cost, and density.

\item 3D stacking also needs unique power and thermal solutions, as traditional heat sink technology would not suffice if we want to add more computation close to the memory. Most of the proposed architectures do not take into account the strict power budget of 3D stacked memories, which limits the architecture's practicality.

\item DRAM and NVM have different memory attributes. A hybrid design can revolutionize our current systems. Huang et al.~\cite{qian2015study} evaluated a 3D heterogeneous storage structure that tightly integrates CPU, DRAM, and a flash-based NVM to meet the memory needs of big data applications, i.e., larger capacity, smaller delay, and wider bandwidth. Processing near heterogeneous memories is a new research topic with high potential (could provide the best of both worlds), and in the future, we expect much interest in this direction.

\item Most of the evaluated architectures focus on the compute aspect. Few architectures focus on providing coherency and virtual memory support. As highlighted in Section~\ref{sec:interondpcachevirtual}, lack of coherency and virtual support makes programming difficult and obstructs the adoption of this paradigm. 

\item More quantitative exploration is required for interconnect networks between the near-memory compute units and also between the host and near-memory compute system. The interplay of NMC units with the emerging interconnect standards like GenZ~\cite{genZdram}, CXL~\cite{cxlwhitepaper}, CCIX~\cite{ccixwhitepaper} and OpenCAPI~\cite{stuecheli2018ibm} could be vital in improving the performance and energy efficiency of big data workloads running on NMC enabled servers.

\item At the application level, algorithms need to provide code and data co-location for energy efficient processing. For example, in case of HMC algorithms should avoid excessive movement of data between vaults and across different memory modules. Whenever it is not possible to avoid an inter-vault data transfer, lightweight mechanisms for data migration should be provided.

\item The field requires a generic set of open-source tools and techniques for these novel systems, as often researchers have to spend a significant amount of time and effort in building the needed simulation environment. Application characterization tools should add support for static and dynamic decision support for offloading processing and data to near-memory systems. To assist in the offloading decision, new metrics are required that could assess whether an application is suitable for these architectures. These tools could provide region-of-interest (or hotspots) in an application that should be offloaded to an NMC system. Besides, a standard benchmark set is missing to gauge different architectural proposals in this domain.

\end{itemize}

\section{Conclusion}
The decline in performance and energy consumption for emerging big-data workloads on the conventional systems, has stimulated an enormous amount of research in processing close to memory. However, to embrace this paradigm, we need to provide a complete ecosystem from hardware to software stack. In this paper, we have analyzed and organized the extensive literature of placing compute units close to the memory using different synonyms (e.g., \textit{processing-in memory}, \textit{near-data processing}) under the umbrella of \textit{near-memory computing} (NMC). This organization is done to distinguish it from the in-situ \textit{computation-in-memory} through novel non-volatile memories such as memristors and phase-change memory. By systematically reviewing the existing NMC systems, we have identified key challenges that need to be addressed in the future. We stress the demand for sophisticated tools and techniques to enable the design space exploration for these novel architectures. Also, we have described our approach to address a subset of those challenges.


\section*{Acknowledgment}
\label{sec:acknowledgment}
This work was performed in the framework of Horizon 2020 program for the project ``Near-Memory Computing (NeMeCo)" and is funded by European Commission under Marie Sklodowska-Curie Innovative Training Networks European Industrial Doctorate (Project ID: 676240).

\ifCLASSOPTIONcaptionsoff
  \newpage
\fi

\bibliographystyle{IEEEtran}
\bibliography{IEEEabrv,ref}

\begin{thebibliography}{100}
\providecommand{\url}[1]{#1}
\csname url@samestyle\endcsname
\providecommand{\newblock}{\relax}
\providecommand{\bibinfo}[2]{#2}
\providecommand{\BIBentrySTDinterwordspacing}{\spaceskip=0pt\relax}
\providecommand{\BIBentryALTinterwordstretchfactor}{4}
\providecommand{\BIBentryALTinterwordspacing}{\spaceskip=\fontdimen2\font plus
\BIBentryALTinterwordstretchfactor\fontdimen3\font minus
  \fontdimen4\font\relax}
\providecommand{\BIBforeignlanguage}[2]{{%
\expandafter\ifx\csname l@#1\endcsname\relax
\typeout{** WARNING: IEEEtran.bst: No hyphenation pattern has been}%
\typeout{** loaded for the language `#1'. Using the pattern for}%
\typeout{** the default language instead.}%
\else
\language=\csname l@#1\endcsname
\fi
#2}}
\providecommand{\BIBdecl}{\relax}
\BIBdecl

\bibitem{Wulf1995}
W.~A. Wulf and S.~A. McKee, ``{Hitting the Memory Wall: Implications of the
  Obvious},'' \emph{SIGARCH Comput. Archit. News}, vol.~23, no.~1, pp. 20--24,
  Mar. 1995.

\bibitem{1050511}
R.~H. Dennard, F.~H. Gaensslen, V.~L. Rideout, E.~Bassous, and A.~R. LeBlanc,
  ``{Design of Ion-implanted {MOSFET}'s with Very Small Physical Dimensions},''
  \emph{IEEE Journal of Solid-State Circuits}, vol.~9, no.~5, pp. 256--268, Oct
  1974.

\bibitem{6175879}
H.~Esmaeilzadeh, E.~Blem, R.~S. Amant, K.~Sankaralingam, and D.~Burger, ``{Dark
  Silicon and The End of Multicore Scaling},'' \emph{IEEE Micro}, vol.~32,
  no.~3, pp. 122--134, 2012.

\bibitem{nair2015active}
R.~{Nair}, S.~F. {Antao}, C.~{Bertolli}, P.~{Bose}, J.~R. {Brunheroto},
  T.~{Chen}, C.~. {Cher}, C.~H.~A. {Costa}, J.~{Doi}, C.~{Evangelinos}, B.~M.
  {Fleischer}, T.~W. {Fox}, D.~S. {Gallo}, L.~{Grinberg}, J.~A. {Gunnels},
  A.~C. {Jacob}, P.~{Jacob}, H.~M. {Jacobson}, T.~{Karkhanis}, C.~{Kim}, J.~H.
  {Moreno}, J.~K. {O'Brien}, M.~{Ohmacht}, Y.~{Park}, D.~A. {Prener}, B.~S.
  {Rosenburg}, K.~D. {Ryu}, O.~{Sallenave}, M.~J. {Serrano}, P.~D.~M. {Siegl},
  K.~{Sugavanam}, and Z.~{Sura}, ``{Active Memory Cube: A Processing-in-Memory
  Architecture for Exascale Systems},'' \emph{IBM Journal of Research and
  Development}, vol.~59, no. 2/3, pp. 17:1--17:14, March 2015.

\bibitem{6898703}
R.~Jongerius, S.~Wijnholds, R.~Nijboer, and H.~Corporaal, ``{An End-to-End
  Computing Model for the Square Kilometre Array},'' \emph{Computer}, vol.~47,
  no.~9, pp. 48--54, Sept 2014.

\bibitem{awan2016micro}
A.~J. Awan, M.~Brorsson, V.~Vlassov, and E.~Ayguade, ``{Micro-Architectural
  Characterization of Apache Spark on Batch and Stream Processing Workloads},''
  in \emph{Big Data and Cloud Computing (BDCloud), Social Computing and
  Networking (SocialCom), Sustainable Computing and Communications
  (SustainCom)(BDCloud-SocialCom-SustainCom), 2016 IEEE International
  Conferences on}.\hskip 1em plus 0.5em minus 0.4em\relax IEEE, 2016, pp.
  59--66.

\bibitem{javed2015performance}
------, ``{Performance Characterization of In-Memory Data Analytics on a Modern
  Cloud Server},'' in \emph{Big Data and Cloud Computing (BDCloud), 2015 IEEE
  Fifth International Conference on}.\hskip 1em plus 0.5em minus 0.4em\relax
  IEEE, 2015, pp. 1--8.

\bibitem{awan2016node}
{Awan, Ahsan Javed and Brorsson, Mats and Vlassov, Vladimir and Ayguade,
  Eduard}, ``{Node Architecture Implications for In-Memory Data Analytics on
  Scale-in Clusters},'' in \emph{Big Data Computing Applications and
  Technologies (BDCAT), 2016 IEEE/ACM 3rd International Conference on}.\hskip
  1em plus 0.5em minus 0.4em\relax IEEE, 2016, pp. 237--246.

\bibitem{6242474}
J.~Jeddeloh and B.~Keeth, ``{Hybrid Memory Cube New DRAM Architecture Increases
  Density and Performance},'' in \emph{2012 Symposium on VLSI Technology
  (VLSIT)}, June 2012, pp. 87--88.

\bibitem{7477494}
J.~T. Pawlowski, ``{Hybrid Memory Cube (HMC)},'' in \emph{2011 IEEE Hot Chips
  23 Symposium (HCS)}, Aug 2011, pp. 1--24.

\bibitem{6757501}
D.~U. {Lee}, K.~W. {Kim}, K.~W. {Kim}, H.~{Kim}, J.~Y. {Kim}, Y.~J. {Park},
  J.~H. {Kim}, D.~S. {Kim}, H.~B. {Park}, J.~W. {Shin}, J.~H. {Cho}, K.~H.
  {Kwon}, M.~J. {Kim}, J.~{Lee}, K.~W. {Park}, B.~{Chung}, and S.~{Hong},
  ``{25.2 A 1.2V 8Gb 8-Channel 128GB/s High-Bandwidth Memory (HBM) Stacked DRAM
  with Effective Microbump I/O Test Methods Using 29nm Process and TSV},'' in
  \emph{2014 IEEE International Solid-State Circuits Conference Digest of
  Technical Papers (ISSCC)}, Feb 2014, pp. 432--433.

\bibitem{6025219}
J.~{Kim}, C.~S. {Oh}, H.~{Lee}, D.~{Lee}, H.~R. {Hwang}, S.~{Hwang}, B.~{Na},
  J.~{Moon}, J.~{Kim}, H.~{Park}, J.~{Ryu}, K.~{Park}, S.~K. {Kang}, S.~{Kim},
  H.~{Kim}, J.~{Bang}, H.~{Cho}, M.~{Jang}, C.~{Han}, J.~{LeeLee}, J.~S.
  {Choi}, and Y.~{Jun}, ``{A 1.2 V 12.8 GB/s 2 Gb Mobile Wide-I/O DRAM With
  4$\times$128 I/Os Using TSV Based Stacking},'' \emph{IEEE Journal of
  Solid-State Circuits}, vol.~47, no.~1, pp. 107--116, Jan 2012.

\bibitem{7092668}
S.~Hamdioui, L.~Xie, H.~A.~D. Nguyen, M.~Taouil, K.~Bertels, H.~Corporaal,
  H.~Jiao, F.~Catthoor, D.~Wouters, L.~Eike, and J.~van Lunteren, ``{Memristor
  Based Computation-in-memory Architecture for Data-Intensive Applications},''
  in \emph{2015 Design, Automation Test in Europe Conference Exhibition
  (DATE)}, March 2015, pp. 1718--1725.

\bibitem{7097722}
H.~Zhang, G.~Chen, B.~C. Ooi, K.~L. Tan, and M.~Zhang, ``{In-Memory Big Data
  Management and Processing: A Survey},'' \emph{IEEE Transactions on Knowledge
  and Data Engineering}, vol.~27, no.~7, pp. 1920--1948, July 2015.

\bibitem{Keeton:1998:CID:290593.290602}
K.~Keeton, D.~A. Patterson, and J.~M. Hellerstein, ``{A Case for Intelligent
  Disks (IDISKs)},'' \emph{SIGMOD Rec.}, vol.~27, no.~3, pp. 42--52, Sep. 1998.

\bibitem{Riedel:1998:ASL:645924.671345}
E.~Riedel, G.~A. Gibson, and C.~Faloutsos, ``{Active Storage for Large-Scale
  Data Mining and Multimedia},'' in \emph{Proceedings of the 24rd International
  Conference on Very Large Data Bases}, ser. VLDB '98.\hskip 1em plus 0.5em
  minus 0.4em\relax San Francisco, CA, USA: Morgan Kaufmann Publishers Inc.,
  1998, pp. 62--73.

\bibitem{7151782}
R.~Nair, ``{Evolution of Memory Architecture},'' \emph{Proceedings of the
  IEEE}, vol. 103, no.~8, pp. 1331--1345, Aug 2015.

\bibitem{siegl2016data}
P.~Siegl, R.~Buchty, and M.~Berekovic, ``{Data-Centric Computing Frontiers: A
  Survey On Processing-In-Memory},'' in \emph{Proceedings of the Second
  International Symposium on Memory Systems}.\hskip 1em plus 0.5em minus
  0.4em\relax ACM, 2016, pp. 295--308.

\bibitem{ghose2018enabling}
S.~Ghose, K.~Hsieh, A.~Boroumand, R.~Ausavarungnirun, and O.~Mutlu, ``{Enabling
  the Adoption of Processing-in-Memory: Challenges, Mechanisms, Future Research
  Directions},'' \emph{arXiv preprint arXiv:1802.00320}, 2018.

\bibitem{629e227db47e4e89b539b0e072ecce08}
G.~{Singh}, L.~{Chelini}, S.~{Corda}, A.~{Javed Awan}, S.~{Stuijk},
  R.~{Jordans}, H.~{Corporaal}, and A.~{Boonstra}, ``{A Review of Near-Memory
  Computing Architectures: Opportunities and Challenges},'' in \emph{2018 21st
  Euromicro Conference on Digital System Design (DSD)}, Aug 2018, pp. 608--617.

\bibitem{ALogicinMemoryComputer}
H.~S. Stone, ``{A Logic-in-Memory Computer},'' \emph{IEEE Transactions on
  Computers}, vol. C-19, no.~1, pp. 73--78, Jan 1970.

\bibitem{5727436}
D.~G. Elliott, W.~M. Snelgrove, and M.~Stumm, ``{Computational Ram: A
  Memory-SIMD Hybrid and its Application to DSP},'' in \emph{1992 Proceedings
  of the IEEE Custom Integrated Circuits Conference}, May 1992, pp.
  30.6.1--30.6.4.

\bibitem{Kogge1994}
P.~M. Kogge, ``{EXECUBE-A New Architecture for Scaleable MPPs},'' in
  \emph{Proceedings of the 1994 International Conference on Parallel Processing
  - Volume 01}, ser. ICPP '94.\hskip 1em plus 0.5em minus 0.4em\relax
  Washington, DC, USA: IEEE Computer Society, 1994, pp. 77--84.

\bibitem{fbram1994}
M.~F. Deering, S.~A. Schlapp, and M.~G. Lavelle, ``{FBRAM: A New Form of Memory
  Optimized for 3D Graphics},'' in \emph{Proceedings of the 21st Annual
  Conference on Computer Graphics and Interactive Techniques}, ser. SIGGRAPH
  '94.\hskip 1em plus 0.5em minus 0.4em\relax New York, NY, USA: ACM, 1994, pp.
  167--174.

\bibitem{375174}
M.~Gokhale, B.~Holmes, and K.~Iobst, ``{Processing in Memory: The Terasys
  Massively Parallel PIM Array},'' \emph{Computer}, vol.~28, no.~4, pp. 23--31,
  Apr 1995.

\bibitem{592312}
D.~Patterson, T.~Anderson, N.~Cardwell, R.~Fromm, K.~Keeton, C.~Kozyrakis,
  R.~Thomas, and K.~Yelick, ``{A Case for Intelligent RAM},'' \emph{IEEE
  Micro}, vol.~17, no.~2, pp. 34--44, Mar 1997.

\bibitem{808425}
Y.~Kang, W.~Huang, S.-M. Yoo, D.~Keen, Z.~Ge, V.~Lam, P.~Pattnaik, and
  J.~Torrellas, ``{FlexRAM: Toward an Advanced Intelligent Memory System},'' in
  \emph{Proceedings 1999 IEEE International Conference on Computer Design: VLSI
  in Computers and Processors (Cat. No.99CB37040)}, 1999, pp. 192--201.

\bibitem{612252}
C.~E. Kozyrakis, S.~Perissakis, D.~Patterson, T.~Anderson, K.~Asanovic,
  N.~Cardwell, R.~Fromm, J.~Golbus, B.~Gribstad, K.~Keeton, R.~Thomas,
  N.~Treuhaft, and K.~Yelick, ``{Scalable Processors in the Billion-Transistor
  Era: IRAM},'' \emph{Computer}, vol.~30, no.~9, pp. 75--78, Sep 1997.

\bibitem{7284059}
J.~Ahn, S.~Hong, S.~Yoo, O.~Mutlu, and K.~Choi, ``{A Scalable
  Processing-in-Memory Accelerator for Parallel Graph Processing},'' in
  \emph{2015 ACM/IEEE 42nd Annual International Symposium on Computer
  Architecture (ISCA)}, June 2015, pp. 105--117.

\bibitem{hsieh2016accelerating}
K.~Hsieh, S.~Khan, N.~Vijaykumar, K.~K. Chang, A.~Boroumand, S.~Ghose, and
  O.~Mutlu, ``{Accelerating Pointer Chasing in 3D-Stacked Memory: Challenges,
  Mechanisms, Evaluation},'' in \emph{2016 IEEE 34th International Conference
  on Computer Design (ICCD)}.\hskip 1em plus 0.5em minus 0.4em\relax IEEE,
  2016, pp. 25--32.

\bibitem{Azarkhish:2016:DEP:2963802.2963805}
E.~Azarkhish, D.~Rossi, I.~Loi, and L.~Benini, ``{Design and Evaluation of a
  Processing-in-Memory Architecture for the Smart Memory Cube},'' in
  \emph{Proceedings of the 29th International Conference on Architecture of
  Computing Systems -- ARCS 2016 - Volume 9637}.\hskip 1em plus 0.5em minus
  0.4em\relax New York, NY, USA: Springer-Verlag New York, Inc., 2016, pp.
  19--31.

\bibitem{7927081}
P.~C. Santos, G.~F. Oliveira, D.~G. Tom{\'e}, M.~A. Alves, E.~C. Almeida, and
  L.~Carro, ``{Operand Size Reconfiguration for Big Data Processing in
  Memory},'' in \emph{Proceedings of the Conference on Design, Automation \&
  Test in Europe}.\hskip 1em plus 0.5em minus 0.4em\relax European Design and
  Automation Association, 2017, pp. 710--715.

\bibitem{loh2013processing}
G.~Loh, N.~Jayasena, M.~Oskin, M.~Nutter, D.~Roberts, M.~Meswani, D.~Zhang, and
  M.~Ignatowski, ``{A Processing in Memory Taxonomy and a Case for Studying
  Fixed-Function PIM},'' in \emph{Workshop on Near-Data Processing (WoNDP)},
  2013.

\bibitem{cho2013xsd}
B.~Y. Cho, W.~S. Jeong, D.~Oh, and W.~W. Ro, ``{XSD: Accelerating MapReduce by
  Harnessing the GPU inside an SSD},'' in \emph{Proceedings of the 1st Workshop
  on Near-Data Processing}, 2013.

\bibitem{kang2013enabling}
Y.~Kang, Y.-s. Kee, E.~L. Miller, and C.~Park, ``{Enabling Cost-effective Data
  Processing with Smart SSD},'' in \emph{Mass Storage Systems and Technologies
  (MSST), 2013 IEEE 29th Symposium on}.\hskip 1em plus 0.5em minus 0.4em\relax
  IEEE, 2013, pp. 1--12.

\bibitem{seshadri2014willow}
\BIBentryALTinterwordspacing
S.~Seshadri, M.~Gahagan, S.~Bhaskaran, T.~Bunker, A.~De, Y.~Jin, Y.~Liu, and
  S.~Swanson, ``{Willow: A User-programmable SSD},'' in \emph{Proceedings of
  the 11th USENIX Conference on Operating Systems Design and Implementation},
  ser. OSDI'14.\hskip 1em plus 0.5em minus 0.4em\relax Berkeley, CA, USA:
  USENIX Association, 2014, pp. 67--80. [Online]. Available:
  \url{http://dl.acm.org/citation.cfm?id=2685048.2685055}
\BIBentrySTDinterwordspacing

\bibitem{6844483}
S.~H. Pugsley, J.~Jestes, H.~Zhang, R.~Balasubramonian, V.~Srinivasan,
  A.~Buyuktosunoglu, A.~Davis, and F.~Li, ``{NDC: Analyzing the Impact of
  3D-Stacked Memory+Logic Devices on MapReduce Workloads},'' in \emph{2014 IEEE
  International Symposium on Performance Analysis of Systems and Software
  (ISPASS)}, March 2014, pp. 190--200.

\bibitem{zhang2014top}
D.~Zhang, N.~Jayasena, A.~Lyashevsky, J.~L. Greathouse, L.~Xu, and
  M.~Ignatowski, ``{TOP-PIM: Throughput-Oriented Programmable Processing in
  Memory},'' in \emph{Proceedings of the 23rd international symposium on
  High-performance parallel and distributed computing}.\hskip 1em plus 0.5em
  minus 0.4em\relax ACM, 2014, pp. 85--98.

\bibitem{xi2015beyond}
S.~L. Xi, O.~Babarinsa, M.~Athanassoulis, and S.~Idreos, ``{Beyond the Wall:
  Near-Data Processing for Databases},'' in \emph{Proceedings of the 11th
  International Workshop on Data Management on New Hardware}.\hskip 1em plus
  0.5em minus 0.4em\relax ACM, 2015, p.~2.

\bibitem{gokhale2015near}
M.~Gokhale, S.~Lloyd, and C.~Hajas, ``{Near Memory Data Structure
  Rearrangement},'' in \emph{Proceedings of the 2015 International Symposium on
  Memory Systems}.\hskip 1em plus 0.5em minus 0.4em\relax ACM, 2015, pp.
  283--290.

\bibitem{7446059}
M.~Gao and C.~Kozyrakis, ``{HRL: Efficient and Flexible Reconfigurable Logic
  for Near-Data Processing},'' in \emph{2016 IEEE International Symposium on
  High Performance Computer Architecture (HPCA)}, March 2016, pp. 126--137.

\bibitem{wang2015propram}
Y.~Wang, Y.~Han, L.~Zhang, H.~Li, and X.~Li, ``{ProPRAM: Exploiting the
  Transparent Logic Resources in Non-Volatile Memory for Near Data
  Computing},'' in \emph{Proceedings of the 52nd Annual Design Automation
  Conference}.\hskip 1em plus 0.5em minus 0.4em\relax ACM, 2015, p.~47.

\bibitem{Jun:2015:BAB:2872887.2750412}
S.~{Jun}, M.~{Liu}, S.~{Lee}, J.~{Hicks}, J.~{Ankcorn}, M.~{King}, S.~{Xu}, and
  {Arvind}, ``{BlueDBM: An Appliance for Big Data analytics},'' in \emph{2015
  ACM/IEEE 42nd Annual International Symposium on Computer Architecture
  (ISCA)}, June 2015, pp. 1--13.

\bibitem{7056040}
A.~Farmahini-Farahani, J.~H. Ahn, K.~Morrow, and N.~S. Kim, ``{NDA: Near-DRAM
  Acceleration Architecture Leveraging Commodity DRAM Devices and Standard
  Memory Modules},'' in \emph{2015 IEEE 21st International Symposium on High
  Performance Computer Architecture (HPCA)}, Feb 2015, pp. 283--295.

\bibitem{ahn2015pim}
J.~Ahn, S.~Yoo, O.~Mutlu, and K.~Choi, ``{PIM-Enabled Instructions: A
  Low-Overhead, Locality-Aware Processing-in-Memory Architecture},'' in
  \emph{Proceedings of the 42nd Annual International Symposium on Computer
  Architecture}.\hskip 1em plus 0.5em minus 0.4em\relax ACM, 2015, pp.
  336--348.

\bibitem{7551394}
K.~Hsieh, E.~Ebrahimi, G.~Kim, N.~Chatterjee, M.~O'Connor, N.~Vijaykumar,
  O.~Mutlu, and S.~W. Keckler, ``{Transparent Offloading and Mapping (TOM):
  Enabling Programmer-Transparent Near-Data Processing in GPU Systems},'' in
  \emph{ACM SIGARCH Computer Architecture News}, vol.~44, no.~3.\hskip 1em plus
  0.5em minus 0.4em\relax IEEE Press, 2016, pp. 204--216.

\bibitem{gu2016biscuit}
\BIBentryALTinterwordspacing
B.~Gu, A.~S. Yoon, D.-H. Bae, I.~Jo, J.~Lee, J.~Yoon, J.-U. Kang, M.~Kwon,
  C.~Yoon, S.~Cho, J.~Jeong, and D.~Chang, ``{Biscuit: A Framework for
  Near-data Processing of Big Data Workloads},'' \emph{SIGARCH Comput. Archit.
  News}, vol.~44, no.~3, pp. 153--165, Jun. 2016. [Online]. Available:
  \url{http://doi.acm.org/10.1145/3007787.3001154}
\BIBentrySTDinterwordspacing

\bibitem{7756764}
A.~Pattnaik, X.~Tang, A.~Jog, O.~Kayiran, A.~K. Mishra, M.~T. Kandemir,
  O.~Mutlu, and C.~R. Das, ``{Scheduling Techniques for GPU Architectures with
  Processing-in-Memory Capabilities},'' in \emph{2016 International Conference
  on Parallel Architecture and Compilation Techniques (PACT)}, Sept 2016, pp.
  31--44.

\bibitem{istvan2017caribou}
Z.~Istv{\'a}n, D.~Sidler, and G.~Alonso, ``{Caribou: Intelligent Distributed
  Storage},'' \emph{Proceedings of the VLDB Endowment}, vol.~10, no.~11, pp.
  1202--1213, 2017.

\bibitem{vermij2017sorting}
\BIBentryALTinterwordspacing
E.~Vermij, L.~Fiorin, C.~Hagleitner, and K.~Bertels, ``{Sorting Big Data on
  Heterogeneous Near-data Processing Systems},'' in \emph{Proceedings of the
  Computing Frontiers Conference}, ser. CF'17.\hskip 1em plus 0.5em minus
  0.4em\relax New York, NY, USA: ACM, 2017, pp. 349--354. [Online]. Available:
  \url{http://doi.acm.org/10.1145/3075564.3078885}
\BIBentrySTDinterwordspacing

\bibitem{koo2017summarizer}
\BIBentryALTinterwordspacing
G.~Koo, K.~K. Matam, T.~I, H.~V. K.~G. Narra, J.~Li, H.-W. Tseng, S.~Swanson,
  and M.~Annavaram, ``{Summarizer: Trading Communication with Computing Near
  Storage},'' in \emph{Proceedings of the 50th Annual IEEE/ACM International
  Symposium on Microarchitecture}, ser. MICRO-50 '17.\hskip 1em plus 0.5em
  minus 0.4em\relax New York, NY, USA: ACM, 2017, pp. 219--231. [Online].
  Available: \url{http://doi.acm.org/10.1145/3123939.3124553}
\BIBentrySTDinterwordspacing

\bibitem{de2017mondrian}
D.~L. De~Oliveira, M.~Paulo, A.~Daglis, N.~Mirzadeh, D.~Ustiugov,
  J.~Picorel~Obando, B.~Falsafi, B.~Grot, and D.~Pnevmatikatos, ``The
  {M}ondrian {D}ata {E}ngine,'' in \emph{Proceedings of the 44th International
  Symposium on Computer Architecture}, no. EPFL-CONF-227947, 2017.

\bibitem{nai2017graphpim}
L.~{Nai}, R.~{Hadidi}, J.~{Sim}, H.~{Kim}, P.~{Kumar}, and H.~{Kim},
  ``{GraphPIM: Enabling Instruction-Level PIM Offloading in Graph Computing
  Frameworks},'' in \emph{2017 IEEE International Symposium on High Performance
  Computer Architecture (HPCA)}, Feb 2017, pp. 457--468.

\bibitem{alian2018application}
M.~{Alian}, S.~W. {Min}, H.~{Asgharimoghaddam}, A.~{Dhar}, D.~K. {Wang},
  T.~{Roewer}, A.~{McPadden}, O.~{O'Halloran}, D.~{Chen}, J.~{Xiong}, D.~{Kim},
  W.~{Hwu}, and N.~S. {Kim}, ``{Application-Transparent Near-Memory Processing
  Architecture with Memory Channel Network},'' in \emph{2018 51st Annual
  IEEE/ACM International Symposium on Microarchitecture (MICRO)}, Oct 2018, pp.
  802--814.

\bibitem{lee2018application}
V.~T. Lee, A.~Mazumdar, C.~C. del Mundo, A.~Alaghi, L.~Ceze, and M.~Oskin,
  ``{Application Codesign of Near-Data Processing for Similarity Search},'' in
  \emph{2018 IEEE International Parallel and Distributed Processing Symposium
  (IPDPS)}.\hskip 1em plus 0.5em minus 0.4em\relax IEEE, 2018, pp. 896--907.

\bibitem{liu2018processing}
J.~Liu, H.~Zhao, M.~A. Ogleari, D.~Li, and J.~Zhao, ``{Processing-in-Memory for
  Energy-efficient Neural Network Training: A Heterogeneous Approach},'' in
  \emph{2018 51st Annual IEEE/ACM International Symposium on Microarchitecture
  (MICRO)}.\hskip 1em plus 0.5em minus 0.4em\relax IEEE, 2018, pp. 655--668.

\bibitem{Boroumand:2018:GWC:3296957.3173177}
\BIBentryALTinterwordspacing
A.~Boroumand, S.~Ghose, Y.~Kim, R.~Ausavarungnirun, E.~Shiu, R.~Thakur, D.~Kim,
  A.~Kuusela, A.~Knies, P.~Ranganathan, and O.~Mutlu, ``{Google Workloads for
  Consumer Devices: Mitigating Data Movement Bottlenecks},'' \emph{SIGPLAN
  Not.}, vol.~53, no.~2, pp. 316--331, Mar. 2018. [Online]. Available:
  \url{http://doi.acm.org/10.1145/3296957.3173177}
\BIBentrySTDinterwordspacing

\bibitem{torabzadehkashi2018compstor}
M.~Torabzadehkashi, S.~Rezaei, V.~Alves, and N.~Bagherzadeh, ``{CompStor: An
  In-storage Computation Platform for Scalable Distributed Processing},'' in
  \emph{2018 IEEE International Parallel and Distributed Processing Symposium
  Workshops (IPDPSW)}.\hskip 1em plus 0.5em minus 0.4em\relax IEEE, 2018, pp.
  1260--1267.

\bibitem{stuecheli2018ibm}
J.~Stuecheli, W.~J. Starke, J.~D. Irish, L.~B. Arimilli, D.~Dreps, B.~Blaner,
  C.~Wollbrink, and B.~Allison, ``{IBM POWER9 Opens up a New Era of
  Acceleration Enablement: OpenCAPI},'' vol.~62, no. 4/5.\hskip 1em plus 0.5em
  minus 0.4em\relax IBM, 2018, pp. 8--1.

\bibitem{Qureshi:2009:SHP:1555815.1555760}
M.~K. Qureshi, V.~Srinivasan, and J.~A. Rivers, ``{Scalable High Performance
  Main Memory System Using Phase-Change Memory Technology},'' \emph{SIGARCH
  Comput. Archit. News}, vol.~37, no.~3, pp. 24--33, Jun. 2009.

\bibitem{1609379}
M.~{Hosomi}, H.~{Yamagishi}, T.~{Yamamoto}, K.~{Bessho}, Y.~{Higo},
  K.~{Yamane}, H.~{Yamada}, M.~{Shoji}, H.~{Hachino}, C.~{Fukumoto},
  H.~{Nagao}, and H.~{Kano}, ``{A Novel Nonvolatile Memory with Spin Torque
  Transfer Magnetization Switching: Spin-Ram},'' in \emph{IEEE
  InternationalElectron Devices Meeting, 2005. IEDM Technical Digest.}, Dec
  2005, pp. 459--462.

\bibitem{ranganathan2011microprocessors}
P.~Ranganathan, ``{From Microprocessors to Nanostores: Rethinking Data-Centric
  Systems},'' \emph{Computer}, vol.~44, no.~01, pp. 39--48, jan 2011.

\bibitem{quero2015self}
L.~C. Quero, Y.-S. Lee, and J.-S. Kim, ``{Self-sorting SSD: Producing Sorted
  Data Inside Active SSDs},'' in \emph{Mass Storage Systems and Technologies
  (MSST), 2015 31st Symposium on}.\hskip 1em plus 0.5em minus 0.4em\relax IEEE,
  2015, pp. 1--7.

\bibitem{torabzadehkashi2019catalina}
M.~Torabzadehkashi, S.~Rezaei, A.~Heydarigorji, H.~Bobarshad, V.~Alves, and
  N.~Bagherzadeh, ``{Catalina: In-Storage Processing Acceleration for Scalable
  Big Data Analytics},'' in \emph{2019 27th Euromicro International Conference
  on Parallel, Distributed and Network-Based Processing (PDP)}.\hskip 1em plus
  0.5em minus 0.4em\relax IEEE, 2019, pp. 430--437.

\bibitem{park2014query}
K.~Park, Y.-S. Kee, J.~M. Patel, J.~Do, C.~Park, and D.~J. Dewitt, ``{Query
  Processing on Smart SSDs},'' \emph{IEEE Data Eng. Bull.}, vol.~37, no.~2, pp.
  19--26, 2014.

\bibitem{jun2018grafboost}
S.~{Jun}, A.~{Wright}, S.~{Zhang}, S.~{Xu}, and {Arvind}, ``{GraFBoost: Using
  Accelerated Flash Storage for External Graph Analytics},'' in \emph{2018
  ACM/IEEE 45th Annual International Symposium on Computer Architecture
  (ISCA)}, June 2018, pp. 411--424.

\bibitem{awan2017performance}
A.~J. Awan, ``Performance characterization and optimization of in-memory data
  analytics on a scale-up server,'' Ph.D. dissertation, KTH Royal Institute of
  Technology and Universitat Polit{\`e}cnica de Catalunya, 2017.

\bibitem{Hadidi:2017:CCT:3154814.3155287}
\BIBentryALTinterwordspacing
R.~Hadidi, L.~Nai, H.~Kim, and H.~Kim, ``{CAIRO: A Compiler-Assisted Technique
  for Enabling Instruction-Level Offloading of Processing-In-Memory},''
  \emph{ACM Trans. Archit. Code Optim.}, vol.~14, no.~4, pp. 48:1--48:25, Dec.
  2017. [Online]. Available: \url{http://doi.acm.org/10.1145/3155287}
\BIBentrySTDinterwordspacing

\bibitem{awan2017identifying}
A.~J. Awan \emph{et~al.}, ``{Identifying the Potential of Near Data Processing
  for Apache Spark},'' in \emph{Proceedings of the International Symposium on
  Memory Systems, {MEMSYS} 2017, Alexandria, VA, USA, October 02 - 05, 2017},
  2017, pp. 60--67.

\bibitem{Vtune}
{Intel Vtune Amplifier XE} 2013.
  \url{http://software.intel.com/en-us/node/544393}.

\bibitem{7429299}
M.~Gao, G.~Ayers, and C.~Kozyrakis, ``{Practical Near-Data Processing for
  In-Memory Analytics Frameworks},'' in \emph{2015 International Conference on
  Parallel Architecture and Compilation (PACT)}, Oct 2015, pp. 113--124.

\bibitem{Sura:2015:DAO:2742854.2742863}
\BIBentryALTinterwordspacing
Z.~Sura, A.~Jacob, T.~Chen, B.~Rosenburg, O.~Sallenave, C.~Bertolli, S.~Antao,
  J.~Brunheroto, Y.~Park, K.~O'Brien, and R.~Nair, ``{Data Access Optimization
  in a Processing-in-memory System},'' in \emph{Proceedings of the 12th ACM
  International Conference on Computing Frontiers}, ser. CF '15.\hskip 1em plus
  0.5em minus 0.4em\relax New York, NY, USA: ACM, 2015, pp. 6:1--6:8. [Online].
  Available: \url{http://doi.acm.org/10.1145/2742854.2742863}
\BIBentrySTDinterwordspacing

\bibitem{Wei05anear-memory}
M.~Wei, M.~Snir, J.~Torrellas, and R.~B. Tremaine, ``{A Near-Memory Processor
  for Vector, Streaming and Bit manipulation Workloads},'' in \emph{In The
  Second Watson Conference on Interaction between Architecture, Circuits, and
  Compilers}, 2005.

\bibitem{picorel2017near}
J.~Picorel, D.~Jevdjic, and B.~Falsafi, ``{Near-Memory Address Translation},''
  in \emph{Parallel Architectures and Compilation Techniques (PACT), 2017 26th
  International Conference on}.\hskip 1em plus 0.5em minus 0.4em\relax Ieee,
  2017, pp. 303--317.

\bibitem{7459501}
J.~Lee, J.~H. Ahn, and K.~Choi, ``{Buffered Compares: Excavating the Hidden
  Parallelism Inside DRAM Architectures with Lightweight Logic},'' in
  \emph{2016 Design, Automation Test in Europe Conference Exhibition (DATE)},
  March 2016, pp. 1243--1248.

\bibitem{7459537}
\BIBentryALTinterwordspacing
{Yitbarek, Salessawi Ferede and Yang, Tao and Das, Reetuparna and Austin,
  Todd}, ``{Exploring Specialized Near-memory Processing for Data Intensive
  Operations},'' in \emph{Proceedings of the 2016 Conference on Design,
  Automation \& Test in Europe}, ser. DATE '16.\hskip 1em plus 0.5em minus
  0.4em\relax San Jose, CA, USA: EDA Consortium, 2016, pp. 1449--1452.
  [Online]. Available: \url{http://dl.acm.org/citation.cfm?id=2971808.2972145}
\BIBentrySTDinterwordspacing

\bibitem{Basu:2013:EVM:2508148.2485943}
A.~Basu, J.~Gandhi, J.~Chang, M.~D. Hill, and M.~M. Swift, ``{Efficient Virtual
  Memory for Big Memory Servers},'' \emph{SIGARCH Comput. Archit. News},
  vol.~41, no.~3, pp. 237--248, Jun. 2013.

\bibitem{Boroumand:2019:CEC:3307650.3322266}
\BIBentryALTinterwordspacing
A.~Boroumand, S.~Ghose, M.~Patel, H.~Hassan, B.~Lucia, R.~Ausavarungnirun,
  K.~Hsieh, N.~Hajinazar, K.~T. Malladi, H.~Zheng, and O.~Mutlu, ``{CoNDA:
  Efficient Cache Coherence Support for Near-data Accelerators},'' in
  \emph{Proceedings of the 46th International Symposium on Computer
  Architecture}, ser. ISCA '19.\hskip 1em plus 0.5em minus 0.4em\relax New
  York, NY, USA: ACM, 2019, pp. 629--642. [Online]. Available:
  \url{http://doi.acm.org/10.1145/3307650.3322266}
\BIBentrySTDinterwordspacing

\bibitem{xiao2018prometheus}
Y.~Xiao, S.~Nazarian, and P.~Bogdan, ``{Prometheus: Processing-in-Memory
  Heterogeneous Architecture Design from a Multi-Layer Network Theoretic
  Strategy},'' in \emph{2018 Design, Automation \& Test in Europe Conference \&
  Exhibition (DATE)}.\hskip 1em plus 0.5em minus 0.4em\relax IEEE, 2018, pp.
  1387--1392.

\bibitem{6557175}
Y.~S. Shao and D.~Brooks, ``{ISA-Independent Workload Characterization and its
  Implications for Specialized Architectures},'' in \emph{2013 IEEE
  International Symposium on Performance Analysis of Systems and Software
  (ISPASS)}, April 2013, pp. 245--255.

\bibitem{4292057}
K.~Hoste and L.~Eeckhout, ``{Microarchitecture-Independent Workload
  Characterization},'' \emph{IEEE Micro}, vol.~27, no.~3, pp. 63--72, May 2007.

\bibitem{junior2017generic}
G.~F. Oliveira, P.~C. Santos, M.~A. Alves, and L.~Carro, ``{Generic Processing
  in Memory Cycle Accurate Simulator under Hybrid Memory Cube Architecture},''
  in \emph{2017 International Conference on Embedded Computer Systems:
  Architectures, Modeling, and Simulation (SAMOS)}.\hskip 1em plus 0.5em minus
  0.4em\relax IEEE, 2017, pp. 54--61.

\bibitem{NAPEL}
\BIBentryALTinterwordspacing
G.~Singh, J.~G{\'o}mez-Luna, G.~Mariani, G.~F. Oliveira, S.~Corda, S.~Stuijk,
  O.~Mutlu, and H.~Corporaal, ``{NAPEL: Near-Memory Computing Application
  Performance Prediction via Ensemble Learning},'' in \emph{Proceedings of the
  56th Annual Design Automation Conference 2019}, ser. DAC '19.\hskip 1em plus
  0.5em minus 0.4em\relax New York, NY, USA: ACM, 2019, pp. 27:1--27:6.
  [Online]. Available: \url{http://doi.acm.org/10.1145/3316781.3317867}
\BIBentrySTDinterwordspacing

\bibitem{jongerius2017analytic}
R.~Jongerius, A.~Anghel, G.~Dittmann, G.~Mariani, E.~Vermij, and H.~Corporaal,
  ``{Analytic Multi-Core Processor Model for Fast Design-Space Exploration},''
  \emph{IEEE Transactions on Computers}, vol.~67, no.~6, pp. 755--770, 2017.

\bibitem{mirzadeh2015sort}
N.~Mirzadeh, Y.~O. Ko{\c{c}}berber, B.~Falsafi, and B.~Grot, ``{Sort vs. Hash
  Join Revisited for Near-Memory Execution},'' in \emph{5th Workshop on
  Architectures and Systems for Big Data (ASBD 2015)}, no. EPFL-CONF-209121,
  2015.

\bibitem{xu2015scaling}
L.~Xu, D.~P. Zhang, and N.~Jayasena, ``{Scaling Deep Learning on Multiple
  In-Memory Processors},'' in \emph{Proceedings of the 3rd Workshop on
  Near-Data Processing}, 2015.

\bibitem{deLima:2018:DSE:3203217.3203280}
\BIBentryALTinterwordspacing
J.~a. P.~C. de~Lima, P.~C. Santos, M.~A.~Z. Alves, A.~C.~S. Beck, and L.~Carro,
  ``{Design Space Exploration for PIM Architectures in 3D-stacked Memories},''
  in \emph{Proceedings of the 15th ACM International Conference on Computing
  Frontiers}, ser. CF '18.\hskip 1em plus 0.5em minus 0.4em\relax New York, NY,
  USA: ACM, 2018, pp. 113--120. [Online]. Available:
  \url{http://doi.acm.org/10.1145/3203217.3203280}
\BIBentrySTDinterwordspacing

\bibitem{7336224}
M.~A.~Z. Alves, C.~Villavieja, M.~Diener, F.~B. Moreira, and P.~O.~A. Navaux,
  ``{Si{NUCA}: A Validated Micro-Architecture Simulator},'' in \emph{2015 IEEE
  17th International Conference on High Performance Computing and
  Communications, 2015 IEEE 7th International Symposium on Cyberspace Safety
  and Security, and 2015 IEEE 12th International Conference on Embedded
  Software and Systems}, Aug 2015, pp. 605--610.

\bibitem{7529923}
J.~D. Leidel and Y.~Chen, ``{HMC-Sim-2.0: A Simulation Platform for Exploring
  Custom Memory Cube Operations},'' in \emph{2016 IEEE International Parallel
  and Distributed Processing Symposium Workshops (IPDPSW)}, May 2016, pp.
  621--630.

\bibitem{7544479}
D.~I. Jeon and K.~S. Chung, ``{CasHMC: A Cycle-Accurate Simulator for Hybrid
  Memory Cube},'' \emph{IEEE Computer Architecture Letters}, vol.~16, no.~1,
  pp. 10--13, Jan 2017.

\bibitem{ramulator-pim-repo}
{SAFARI Research Group}, ``{{Ramulator for Processing-in-Memory}},''
  \url{https://github.com/CMU-SAFARI/ramulator-pim/}.

\bibitem{DBLP:journals/corr/ChoeLPKCY16}
H.~Choe, S.~Lee, H.~Nam, S.~Park, S.~Kim, E.-Y. Chung, and S.~Yoon,
  ``{Near-Data Processing for Machine Learning},'' \emph{CoRR}, vol.
  abs/1610.02273, 2016.

\bibitem{jo2016data}
Y.-Y. Jo, S.-W. Kim, M.~Chung, and H.~Oh, ``{Data Mining in Intelligent SSD:
  Simulation-Based Evaluation},'' in \emph{2016 International Conference on Big
  Data and Smart Computing (BigComp)}.\hskip 1em plus 0.5em minus 0.4em\relax
  IEEE, 2016, pp. 123--128.

\bibitem{Binkert:2011:GS:2024716.2024718}
\BIBentryALTinterwordspacing
N.~Binkert, B.~Beckmann, G.~Black, S.~K. Reinhardt, A.~Saidi, A.~Basu,
  J.~Hestness, D.~R. Hower, T.~Krishna, S.~Sardashti, R.~Sen, K.~Sewell,
  M.~Shoaib, N.~Vaish, M.~D. Hill, and D.~A. Wood, ``{The Gem5 Simulator},''
  \emph{SIGARCH Comput. Archit. News}, vol.~39, no.~2, pp. 1--7, Aug. 2011.
  [Online]. Available: \url{http://doi.acm.org/10.1145/2024716.2024718}
\BIBentrySTDinterwordspacing

\bibitem{chang2012limits}
J.~Chang, P.~Ranganathan, T.~Mudge, D.~Roberts, M.~A. Shah, and K.~T. Lim, ``{A
  Limits Study of Benefits from Nanostore-Based Future Data-Centric System
  Architectures},'' in \emph{Proceedings of the 9th conference on Computing
  Frontiers}.\hskip 1em plus 0.5em minus 0.4em\relax ACM, 2012, pp. 33--42.

\bibitem{cotson-osr09}
E.~Argollo, A.~Falc\'{o}n, P.~Faraboschi, M.~Monchiero, and D.~Ortega,
  ``{COTSon: Infrastructure for Full System Simulation},'' \emph{SIGOPS Oper.
  Syst. Rev.}, vol.~43, no.~1, pp. 52--61, 2009.

\bibitem{yen2008notary}
L.~Yen, S.~C. Draper, and M.~D. Hill, ``Notary: Hardware techniques to enhance
  signatures,'' in \emph{Proceedings of the 41st annual IEEE/ACM International
  Symposium on Microarchitecture}.\hskip 1em plus 0.5em minus 0.4em\relax IEEE
  Computer Society, 2008, pp. 234--245.

\bibitem{gu2009component}
\BIBentryALTinterwordspacing
X.~Gu, I.~Christopher, T.~Bai, C.~Zhang, and C.~Ding, ``{A Component Model of
  Spatial Locality},'' in \emph{Proceedings of the 2009 International Symposium
  on Memory Management}, ser. ISMM '09.\hskip 1em plus 0.5em minus 0.4em\relax
  New York, NY, USA: ACM, 2009, pp. 99--108. [Online]. Available:
  \url{http://doi.acm.org/10.1145/1542431.1542446}
\BIBentrySTDinterwordspacing

\bibitem{anghel2016instrumentation}
A.~Anghel, L.~M. Vasilescu, G.~Mariani, R.~Jongerius, and G.~Dittmann, ``{An
  Instrumentation Approach for Hardware-agnostic Software Characterization},''
  \emph{International Journal of Parallel Programming}, vol.~44, no.~5, pp.
  924--948, 2016.

\bibitem{pouchet2012polybench}
L.-N. Pouchet, ``{Polybench: The Polyhedral Benchmark Suite},'' \emph{URL:
  http://www. cs. ucla. edu/pouchet/software/polybench}, 2012.

\bibitem{corda2019scopes}
\BIBentryALTinterwordspacing
S.~Corda, G.~Singh, A.~J. Awan, R.~Jordans, and H.~Corporaal, ``{Memory and
  Parallelism Analysis Using a Platform-Independent Approach},'' in
  \emph{Proceedings of the 22nd International Workshop on Software and
  Compilers for Embedded Systems}, ser. SCOPES '19.\hskip 1em plus 0.5em minus
  0.4em\relax New York, NY, USA: ACM, 2019, pp. 23--26. [Online]. Available:
  \url{http://doi.acm.org/10.1145/3323439.3323988}
\BIBentrySTDinterwordspacing

\bibitem{corda2019DSD}
------, ``{Platform Independent Software Analysis for Near Memory Computing},''
  in \emph{"Proceedings of 22nd Euromicro Conference on Digital System Design,
  DSD"}, 2019.

\bibitem{zinenko:hal-01965599}
\BIBentryALTinterwordspacing
O.~Zinenko, L.~Chelini, and T.~Grosser, ``{Declarative Transformations in the
  Polyhedral Model},'' {Inria ; ENS Paris - Ecole Normale Sup{\'e}rieure de
  Paris ; ETH Zurich ; TU Delft ; IBM Z{\"u}rich}, Research Report RR-9243,
  Dec. 2018. [Online]. Available: \url{https://hal.inria.fr/hal-01965599}
\BIBentrySTDinterwordspacing

\bibitem{grosser2012polly}
T.~Grosser, A.~Groesslinger, and C.~Lengauer, ``{Polly-—Performing Polyhedral
  Optimizations on a Low-Level Intermediate Representation},'' \emph{Parallel
  Processing Letters}, vol.~22, no.~04, p. 1250010, 2012.

\bibitem{akin2015data}
B.~Akin, F.~Franchetti, and J.~C. Hoe, ``{Data Reorganization in Memory Using
  3D-Stacked DRAM},'' in \emph{Proceedings of the 42nd Annual International
  Symposium on Computer Architecture}.\hskip 1em plus 0.5em minus 0.4em\relax
  ACM, 2015, pp. 131--143.

\bibitem{7856642}
Q.~{Guo}, T.~{Low}, N.~{Alachiotis}, B.~{Akin}, L.~{Pileggi}, J.~C. {Hoe}, and
  F.~{Franchetti}, ``{Enabling Portable Energy Efficiency with Memory
  Accelerated Library},'' in \emph{2015 48th Annual IEEE/ACM International
  Symposium on Microarchitecture (MICRO)}, Dec 2015, pp. 750--761.

\bibitem{10.1007/978-3-642-15582-6_49}
S.~Verdoolaege, ``{isl: An Integer Set Library for the Polyhedral Model},'' in
  \emph{Mathematical Software -- ICMS 2010}, K.~Fukuda, J.~v.~d. Hoeven,
  M.~Joswig, and N.~Takayama, Eds.\hskip 1em plus 0.5em minus 0.4em\relax
  Berlin, Heidelberg: Springer Berlin Heidelberg, 2010, pp. 299--302.

\bibitem{qian2015study}
C.~Qian, L.~Huang, P.~Xie, N.~Xiao, and Z.~Wang, ``{A Study on Non-Volatile 3D
  Stacked Memory for Big Data Applications},'' in \emph{International
  Conference on Algorithms and Architectures for Parallel Processing}.\hskip
  1em plus 0.5em minus 0.4em\relax Springer, 2015, pp. 103--118.

\bibitem{genZdram}
\BIBentryALTinterwordspacing
M.~Krause and M.~Witkowski, ``{Gen-Z DRAM and Persistent Memory Theory of
  Operation},'' \emph{Gen-Z Consortium White Paper}, 2019. [Online]. Available:
  \url{{http://genzconsortium.org/wp-content/uploads/2019/03/Gen-Z-DRAM-PM-Theory-of-Operation-WP.pdf}}
\BIBentrySTDinterwordspacing

\bibitem{cxlwhitepaper}
\BIBentryALTinterwordspacing
D.~D. Sharma, ``{Compute Express Link},'' \emph{CXL Consortium White Paper}.
  [Online]. Available:
  \url{{https://docs.wixstatic.com/ugd/0c1418_d9878707bbb7427786b70c3c91d5fbd1.pdf}}
\BIBentrySTDinterwordspacing

\bibitem{ccixwhitepaper}
\BIBentryALTinterwordspacing
``{An Introduction to CCIX White Paper},'' \emph{CCIX Consortium Inc}.
  [Online]. Available:
  \url{{https://docs.wixstatic.com/ugd/0c1418_c6d7ec2210ae47f99f58042df0006c3d.pdf}}
\BIBentrySTDinterwordspacing

\end{thebibliography}

\end{document}